\documentclass[a4paper,11pt]{article}
\usepackage{jheppub} 
\usepackage{lineno}
\usepackage{slashed}
\usepackage{braket}

\def\lsim{\mathrel{\rlap{\lower4pt\hbox{\hskip1pt$\sim$}}
  \raise1pt\hbox{$<$}}}
\def\gsim{\mathrel{\rlap{\lower4pt\hbox{\hskip1pt$\sim$}}
  \raise1pt\hbox{$>$}}}
\newcommand{\vev}[1]{ \left\langle {#1} \right\rangle }

\newcommand{\beq}{\begin{equation}}
\newcommand{\eeq}{\end{equation}}
\newcommand{\bea}{\begin{eqnarray}}
\newcommand{\eea}{\end{eqnarray}}
\newcommand{\nn}{\nonumber\\}

\newcommand{\bwt}{\begin{widetext}}
\newcommand{\ewt}{\end{widetext}}

\newcommand{\mbphi}{\overline{m}_\phi}
\newcommand{\mbpsi}{\overline{m}_\psi}

\newcommand{\Mbpsi}{\overline{M}_\psi}

\graphicspath{{./Figures/}}

\newcommand{\fref}[1]{Fig.\,\ref{fig:#1}} 
\newcommand{\eref}[1]{Eq.\,\eqref{Eq:#1}} 
\newcommand{\aref}[1]{Appendix\,\ref{app:#1}}
\newcommand{\sref}[1]{Section~\ref{sec:#1}}

\title{\boldmath Classical constant electric fields and\\ the Schwinger effect in de Sitter}

\author[a]{Mar Bastero-Gil}
\author[a,b]{Paulo B. Ferraz}
\author[b]{António Torres Manso}
\author[c,d]{Lorenzo Ubaldi}
\author[a]{Roberto Vega-Morales}
\affiliation[a]{Departamento de F\'{i}sica Te\'{o}rica y del Cosmos and CAFPE,\\
Universidad de Granada, Campus de Fuentenueva, E-18071 Granada, Spain}
\affiliation[b]{Universidad de Coimbra, Faculdade de Ciências e Tecnologia da Universidade\\
de Coimbra and CFisUC, Rua Larga, 3004-516 Coimbra, Portugal}
\affiliation[c]{Jo\v{z}ef Stefan Institute, Jamova 39, 1000 Ljubljana, Slovenia}
\affiliation[d]{Institute for Fundamental Physics of the Universe (IFPU), \\ Via Beirut 2, 34014 Trieste, Italy}

\emailAdd{mbg@ugr.es}
\emailAdd{paulo.ferraz@student.uc.pt}
\emailAdd{atmanso@uc.pt}
\emailAdd{lorenzo.ubaldi@ijs.si}
\emailAdd{rvegamorales@ugr.es}

\abstract{
We study constant classical electric fields and the Schwinger effect in de Sitter space, with potential implications for magnetogenesis and inflationary dark matter production.~Treating the photon as a dynamical field, we show that sustaining a constant electric field in de Sitter requires a tachyonic photon mass of order the Hubble scale.~This observation has physical implications, as it alters the infrared behaviour of the induced Schwinger current.~Using an on-shell renormalization condition consistent with a tachyonic photon, we recompute the current for charged fermions and scalars, finding it to be finite and positive even in the massless limit of the charge carriers, contrary to earlier results in the literature predicting a puzzling negative IR divergence.~For scalars, we include a non-minimal coupling to the Ricci curvature, enabling us to analyze the conformal limit, where the current closely matches that of charged fermions.
}

\begin{document}
\maketitle
\flushbottom

\section{Introduction}

A strong electric field is expected to generate electron-positron pairs directly from the quantum field theory (QFT) vacuum.~This possibility was first anticipated in 1936 by Euler and Heisenberg~\cite{Heisenberg:1936nmg}, well before the modern formulation of Quantum Electrodynamics (QED), and later firmly established in Schwinger’s seminal work~\cite{Schwinger:1951nm}.~The effect, known as Schwinger pair production, has not yet been observed in a laboratory setting\footnote{Observations of the Schwinger effect have been reported in condensed matter systems~\cite{Berdyugin:2021njg,Schmitt:2022pkd}, where graphene provides an analog of the QFT vacuum~\cite{Allor:2007ei}, and electron–hole pairs replace $e^+e^-$ pairs.}.~This leaves us in the intriguing situation that one of the fundamental predictions of QED, arguably the most successful theory of nature, remains untested experimentally.~The main obstacle is the requirement of an extremely strong electric field, on the order of the electron mass squared, which exceeds current technological capabilities.~Nonetheless, ongoing advances keep alive the prospect of observing this phenomenon in the laboratory in the near future~\cite{Dunne:2008kc,Fedotov:2022ely}.

While such extreme fields remain out of reach in the laboratory, the situation may have been different in the early Universe.~During inflation, the vast energy available could naturally generate electric fields strong enough to trigger the Schwinger effect.~Moreover, as we discuss in detail, in an expanding spacetime the Schwinger effect can occur even in the presence of very weak electric fields.~As explored in numerous studies~\cite{Anber:2006xt,Anber:2009ua,Graham:2015rva,Adshead:2016iae,Bastero-Gil:2018uel,Bastero-Gil:2021wsf,Arvanitaki:2021qlj}, mechanisms for sustaining a nearly constant electric field during inflation can be constructed by coupling a $U(1)$ gauge field to the background expansion.~In this context, the Schwinger effect may play a significant role in both magnetogenesis and the production of dark matter during inflation and raises the intriguing possibility that the first observational evidence of the Schwinger effect could arise from cosmological signals~\cite{Chua:2018dqh}. 
	
The Schwinger current in de Sitter space was first computed in $1+1$ dimensions~\cite{Frob:2014zka,Stahl:2015gaa,Stahl:2016geq}, and later extended to $3+1$ dimensions for minimally coupled charged particles of spin 0~\cite{Kobayashi:2014zza,Bavarsad:2016cxh,Hayashinaka:2016dnt,Banyeres:2018aax,Bavarsad:2017oyv,Stahl:2018idd} and spin 1/2~\cite{Hayashinaka:2016qqn}.~Although the regularized current is formally divergent, it can nevertheless be obtained analytically in a fully non-perturbative manner and made finite through an appropriate renormalization procedure.~In $3+1$ dimensions the renormalized current was found to display some puzzling features and in particular both the scalar~\cite{Kobayashi:2014zza,Banyeres:2018aax} and fermion~\cite{Hayashinaka:2016dnt} currents were found to have a peculiar negative infrared (IR) divergence in the small electron\,\footnote{For the remainder of this paper we use the label ``electron" for both charged scalars and fermions.} mass regime.~This would lead to large currents flowing opposite to the electric field direction and imply a negative conductivity for small electron masses.~It was argued in~\cite{Banyeres:2018aax} that these regimes with divergent IR negative conductivity are not physical and should be interpreted by taking into account the running of the gauge coupling from the small electron mass scale to the Hubble scale.~When the separation between the two scales is very large, one crosses a Landau pole, beyond which the conductivity appears to be negative but the theory is no longer under control.

In this work we argue that the appearance of these IR divergences is due to the renormalization procedures implicitly utilized in~\cite{Kobayashi:2014zza,Banyeres:2018aax,Hayashinaka:2016dnt}.~Imposing physical renormalization conditions which are consistent with a constant electric field in de Sitter space, we show that the renormalized current is free of any IR divergences, both for scalars and fermions, without any need to run the coupling.~In deciding how to fix the renormalization conditions, we first point out that in order to sustain a constant electric field in de Sitter space it is unavoidable to have a tachyonic instability.~Requiring the electric field to be constant and uniform both inside and outside the Hubble horizon, as was assumed in~\cite{Kobayashi:2014zza,Banyeres:2018aax,Hayashinaka:2016dnt}, implies that the photon, treated as a classical dynamical field, must have a tachyonic mass $m_A^2 = -2H^2$, where $H$ is the Hubble constant.~Upon quantization of the electron field, the theory contains a single counterterm to absorb the only divergence corresponding to the wave-function renormalization of the photon field or, equivalently, to the electric charge renormalization.~We fix the counterterm by imposing an on-shell renormalization condition with $m_A^2 = -2H^2$ and show that one obtains a renormalized current {\em free of UV and IR divergences}.~We also show for the first time that for a conformally coupled charged scalar the behavior of the current is the same as for a charged fermion, as one would expect.

This paper is organized as follows:~In \sref{renormlag} we discuss the conditions required for sustaining a classical constant electric field in de Sitter space.~We show that this necessarily implies a tachyonic photon mass and examine the consequences for renormalization, presenting the relevant on-shell renormalization conditions and renormalized Lagrangian.~In \sref{scalarcurrent} we compute the induced Schwinger current for charged scalars in de Sitter space.~We analyze its behavior across different regimes, including the strong and weak-field limits as well as the large mass limit, and compare with previous results obtained in the literature.~In \sref{fermioncurrent} we turn to the fermionic case, performing the analogous computation for the charged fermion Schwinger current, again studying its behavior in various limits.~We summarize our findings and discuss their implications in \sref{conclusion}.~Several technical details are collected in the appendices.

\section{Classical constant electric fields in de Sitter}\label{sec:renormlag}

In this section we discuss a classical constant electric field in de Sitter space.~We are ultimately interested in the production of charged particles in this setup and in the current they generate.~Such particles can be produced both via the Schwinger effect, due to the electric field, and gravitationally, due to the time dependent de Sitter metric.~The current we will compute is defined through the usual Noether's theorem as the variation of the action with respect to the gauge field.~Upon explicit calculation, as we will see, one finds this quantity to diverge.~This is not surprising, as the Schwinger effect is related to loop diagrams in quantum field theory.~The purpose of this work is twofold.~First, we carefully go through the renormalization procedure, for which we already pointed out the main points in a previous work~\cite{Bastero-Gil:2025jio}.~Here we provide more details of the calculation.~Second, we examine our renormalized currents in various limits, which elucidates several important physical aspects and confirms the validity of our approach. 

As in previous studies~\cite{Kobayashi:2014zza, Hayashinaka:2016qqn, Banyeres:2018aax}, we consider a classical constant electric field in de Sitter space and treat the geometry of spacetime as a background.~This means we do not deal with the dynamics of the energy-momentum tensor and do not worry about renormalizing vacuum energy.~Crucially, and in contrast to previous works, we treat the classical gauge field $A_\mu$, which is responsible for setting 
the constant electric field, as \emph{dynamical}.~This is because the only divergence in the calculation of the current is encoded in the vacuum polarization diagram, consisting in an electron loop with two external photon legs.~To remove the divergence we need the counterterm associated with such a diagram.~This requires the two-point function of the photon field, which is obtained by treating the field dynamically.

In this section we point out that in order to have a constant electric field, the dynamics of the gauge field in de Sitter space is not trivial and a tachyonic instability is required.~In later sections we discuss how such a dynamics affects the renormalization condition, which fixes the finite part
of the counterterm, and leads to a well-behaved renormalized current.

\subsection{Conditions for a constant electric field}\label{sec:efield}

We begin with a free abelian gauge theory in a de Sitter background,
\bea
	S=-\int d^4 x \sqrt{-g}\,\frac{1}{4}F^{\mu\nu}F_{\mu\nu}\,.
\eea
Here the metric is $g_{\mu\nu} = a^2(\tau)(-{\rm d}\tau^2 + {\rm d}\vec x^2)$, with the conformal time $\tau$,
\begin{equation}
\tau = - \frac{1}{aH} < 0 \, , \qquad H = \frac{{\rm d}a}{a^2 {\rm d}\tau} = {\rm const.} \, ,
\end{equation}
and $F_{\mu\nu} = \partial_\mu A_\nu - \partial_\nu A_\mu$.~We are interested in the configuration which realizes a constant and uniform electric field.~This is the one measured by a comoving observer with 4-velocity $u^\mu$ ($u^i = 0$, $u_\mu u^\mu = -1$), given by $E_\mu = u^\nu F_{\mu\nu}$ and such that the field strength $E_\mu E^\mu = E^2 = {\rm const}$.~The gauge field configuration that produces such a constant electric field in the $z$ direction is given by~\cite{Kobayashi:2014zza,Hayashinaka:2016qqn,Banyeres:2018aax},
 \bea\label{Eq:Edef}
 A_\mu=\frac{E}{H^2 \tau} \delta_\mu^z \, .
 \eea
Throughout this work, and as done in previous literature, we treat the gauge field as classical.~However, as discussed above and differently from previous literature, we treat it as dynamical, rather than as a background, requiring it to solve the equation of motion,
\bea \label{Eq:EOMA}
g^{\alpha \mu}\partial_\alpha F_{\mu\nu} = 0 \, .
\eea
It is easy to check that \eref{Edef} does not solve \eref{EOMA}:
\bea
g^{\alpha \mu}\partial_\alpha F_{\mu\nu} = - 2 H^2 \tau^2 \frac{E}{H^2 \tau^3} \delta_\nu^z = -2 H^2 A_\nu \neq 0 \, .
\eea
The easiest way to satisfy the equation of motion is to introduce a tachyonic mass,
\bea
m_A^2 = -2H^2 \, ,
\eea
into the action of the gauge field,
\bea\label{Eq:LmA}
	S=-\int d^4 x \sqrt{-g}\,\left(\frac{1}{4}F^{\mu\nu}F_{\mu\nu}+\frac{1}{2} m_A^2 A_\mu A^\mu\right)\,.
\eea
The photon mass $m_A$ should be thought of as a Stueckelberg mass~\cite{Ruegg:2003ps}, obtained upon the replacement
$\frac{1}{2} m_A^2 A_\mu^2 \to \frac{1}{2} m_A^2 \left( A_\mu + \frac{1}{m_A} \partial_\mu \sigma \right)^2$, with $\sigma$ the Stueckelberg field, and the addition of the gauge fixing term ${\cal L}_{\rm gf} = - \sqrt{-g} \frac{1}{2\xi_A} \left( \partial_\mu A^\mu - \xi_A m_A \sigma \right)^2$.~The theory is then gauge invariant.~We see that in order to have a constant electric field in de Sitter space, 
we need a source term which breaks conformal invariance and provides a tachyonic condition, necessary to compensate for the exponential expansion of space.

A second possible solution is to instead introduce a modified kinetic term of the form $(\tau_e/\tau) F_{\mu\nu} F^{\mu\nu}$ where $\tau_e$ is a constant which could be fixed as the time at the end of inflation for example.~As we show in the Appendix~\ref{app:conE}, this leads to the same equations of motion as introducing a tachyonic mass with a canonically normalized kinetic term.~In both scenarios the electric field originates from a transverse component of the gauge field, which we treat as purely classical, with no contribution from the longitudinal mode.~It should not be surprising that in de Sitter space the photon must have a tachyonic mass since, as space expands exponentially, the electric field must be continuously fed exponentially meaning that a tachyonic instability is required to maintain a constant electric field.

A realistic model could be constructed, for example, by considering a 
massless photon with an effective coupling to the inflaton $\phi$ such as an axion like coupling $\phi F_{\mu\nu} \tilde F^{\mu\nu}$.~This acts as a source for the electric field, leads~\cite{Anber:2009ua,Bastero-Gil:2018uel} to tachyonic production of long wavelength modes for one of the two transverse components of the gauge field, and generates a classical  electromagnetic field with $\langle \vec E \cdot \vec B \rangle \neq 0$, coherent over scales comparable to the Hubble horizon.~A full treatment of the Schwinger effect incorporating this mechanism for generating a constant electromagnetic field, including an explicit inflationary model and possible backreaction effects, requires numerical study~\cite{Sobol:2020frh} and left to ongoing work~\cite{followup2}.

In this work, like in previous literature~\cite{Kobayashi:2014zza,Hayashinaka:2016qqn,Banyeres:2018aax} 
directly related the Schwinger current, we concentrate on the simplified assumption that only the zero momentum mode (infinite wavelength) of the gauge field experiences a tachyonic instability, which is provided by the tachyonic mass term in the action.~As a consequence we have only a uniform and constant electric field and no magnetic field.~Although this setup is less realistic than the $\phi F_{\mu\nu} \tilde F^{\mu\nu}$ model, it has the virtue of making the calculation fully tractable {\em analytically} giving useful insight into the Schwinger effect in de Sitter spacetime.~We will see below the crucial role the tachyonic mass condition $m_A^2 = -2H^2$ plays in renormalization and obtaining a physically consistent Schwinger current with proper IR behavior.

\subsection{Renormalized Lagrangian with a constant electric field}\label{sec:renorm}

In the fully quantum theory of QED there are four quantities (ignoring vacuum energy) which need to be renormalized, associated with the normalization of the fields for the photon and electron, the electron mass, and the electric charge.~These four bare parameters receive quantum corrections at loop order which are divergent.~However, since we are treating the electric field as classical, in other words the photon is not quantized, there are no internal photon propagators in the theory at any loop order.~Thus, the electron propagator is not corrected at loop level which implies that neither the electron field or mass are renormalized.~Furthermore, due to the Ward identity, the renormalization of the electric charge is determined by the renormalization of the photon field.~So in the end, when considering a classical constant electric field, there is only one Lagrangian parameter which must be renormalized which we take to be the normalization of the photon field.

Since we are interested in computing the current of charged particles generated by the constant electric field out of the vacuum, one might worry about radiative effects which would require quantizing the photon field.~However, it can be shown~\cite{Bastero-Gil:2023mxm} that after the $e^+ e^-$ pair is produced, the net effect of acceleration due to the electric field and deceleration (redshift) due to the expansion of space is such that within a few e-folds the particles reach terminal velocity which implies that very quickly the current reaches a constant value.~This is a crucial difference between de Sitter and flat spacetime and allows us to treat the gauge field as classical since the charged particles do not accelerate.~Thus we do not need to account for radiative effects and can solve the theory exactly quantizing only the electron field, further justifying our classical treatment of the electric field. 

We can write the renormalized Lagrangian for a classical photon with tachyonic mass $m_A^2 = -2H^2$ in de Sitter coupled to a current as,
\bea\label{Eq:renormlag}
\mathcal{L} &=& 
 - \frac{1}{4} (1 + \delta_3) (F_{\mu \nu})^2 - \frac{1}{2} m_A^2 A_\mu A^\mu 
 - A_\mu J^\mu + ... ~~.
\eea
All fields and parameters are understood to be renormalized, and the conserved current $J^\mu$ can be for fermions or scalars.~One can show~\cite{McKeon:2006ym} that the renormalized $m_A$ is related to the bare one as $m_A^2 = (1- \delta_3)m_{A_0}^2$ in the same way the renormalized charge $e$ is related to the bare charge $e^2 = (1-\delta_3) e^2_0$, so there is no additional divergence.~As we will see, in the presence of a constant classical electric field, the induced Schwinger current only depends on just two ratios:\,the charge times electric field divided by the Hubble scale squared and the electron mass divided by Hubble scale.

The counterterm $\delta_3$ cancels the infinite divergence in the photon field normalization.~In this model the full divergence is encoded in the one-loop vacuum polarization diagram shown in~\fref{selfenergy}.~Note for a scalar there is a second ``bubble" diagram which also contributes.~There are no higher loop corrections to the two point function of the photon, due to the fact that the photon is not quantized.~Thus, once we compute the vacuum polarization diagram(s), and fix $\delta_3$ with an appropriate (physical) renormalization condition, the theory is defined to all orders.~Even non-perturbative quantities such as the induced Schwinger current can be computed with a single counterterm and are guaranteed to be finite and unambiguous predictions of the theory.
\begin{figure}
\vspace{.4cm}
\centering 
\includegraphics[width=.5\columnwidth]{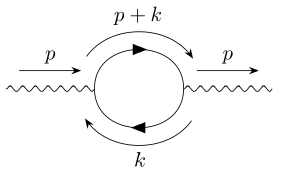}
\caption{Vacuum polarization diagram.}
\label{fig:selfenergy}
\end{figure}

In order to define our theory we must first fix the counterterm with an appropriate \emph{physical} renormalization condition.~Since the field $A_\mu$ is classical, the obvious choice is to pick on-shell renormalization conditions, requiring the renormalized
two point function to have a pole at $p^2 = -m_A^2$ with residue $-i$ which fixes,
\bea
\Pi(p^2 = -m_A^2) = 0, 
\eea
where $e^2\Pi(p^2)$ is defined as the coefficient of $-i\left(p^2 g^{\mu \nu} - p^\mu p^\nu\right)$ in the sum of all 1PI contributions to the photon 2-point correlation function~\cite{Schwartz:2014sze}.~Since we have a classical electric field and $A_\mu$ is not quantized, $\Pi(p^2)$ is determined exactly at one loop, with the corresponding diagram shown in~\fref{selfenergy} (with a second diagram for scalars).~This renormalization condition leads to the condition on the counterterm,
\bea\label{Eq:del3fix}
\delta_3 = - e^2\Pi(-m_A^2) .
\eea
Obviously, for a massless photon one takes $m_A^2 = 0$ which holds in flat space\,\footnote{As we discuss more below, this is equivalent to the renormalization procedures used in~\cite{Kobayashi:2014zza,Hayashinaka:2016qqn,Banyeres:2018aax}.}.~However, as we have discussed above, a massless photon is inconsistent with a constant classical electric field in de Sitter space which requires a tachyonic mass $m_A^2 = -2H^2$.~Thus we take the counterterm to be fixed by the condition,
\bea\label{Eq:del3fix2}
\delta_3 = - e^2\Pi(2H^2) .
\eea
As we'll see, with this renormalization condition, the renormalized Schwinger current in de Sitter space is not only UV finite, but crucially, it is also free of the negative IR divergences which were found in previous calculations for both scalars~\cite{Kobayashi:2014zza,Banyeres:2018aax} and fermions~\cite{Hayashinaka:2016qqn}.~In Appendix~\ref{app:conE} we discuss an alternative way of producing a uniform and constant electric field, but no magnetic field, relying on a modified kinetic term.~We show that at the end of the day the equation of motion for the vector field is the same as in the tachyonic mass case.~That implies that the two cases are physically indistinguishable and that the same renormalization condition described in this section applies to both.

\subsection{Counterterm with a constant electric field}\label{sec:counterterm}

Unlike for the regularized Schwinger current, when computing the vacuum polarization diagram we cannot compute analytically the loop diagram in de Sitter space.~To obtain an analytic result, one must work in the Minkowski limit in both the spacetime integral and the massive charged scalar propagator, as was done (implicitly) in previous calculations in the literature~\cite{Banyeres:2018aax}.~This properly takes care of the UV modes in the counterterm thus allowing divergences in the regularized current to be canceled ensuring a finite renormalized current.~The vacuum polarization diagram is then given by the well known result from QED which depends on the invariant mass squared $(p^2)$ of the photon.~If we follow the renormalization procedure used (implicitly) in previous calculations in the literature, assuming a massless photon and setting $p^2=0$, a negative IR divergence is introduced into the counterterm when the electron mass is taken to zero.~As we'll see, this negative IR divergence is then introduced into the physical renormalized current as there is no corresponding IR divergence in the regularized current which could cancel it.

To see this explicitly we take the flat space result from scalar QED for the vacuum polarization~\cite{Schwartz:2014sze}.~Setting $p^2 = 0$, we have for the counterterm, 
\bea\label{Eq:d3phi1}
\delta_3^{\phi} &=& -e^2\Pi(0) 
=
\left(\frac{e}{12\pi}\right)^2
\Big[ -3 \ln \Big(\frac{\Lambda ^2}{m_\phi^2} \Big) \Big]\, ,
\eea
where $m_\phi$ is the mass of the charged scalar which enters through the propogator (in the Minkowski limit) and $\Lambda$ is the mass of the
Pauli-Villars fields used to regulate the UV divergence (see~\aref{delta3S} for details) which cancels the one in the regularized current.~With this counterterm we can reproduce previous results in the literature for the renormalized current of a minimally coupled massive charged scalar~\cite{Kobayashi:2014zza,Banyeres:2018aax} which were obtained using different renormalization procedures.~While the UV log divergence cancels the one in the regularized current we see explicitly the IR divergence in the $m_\phi \to 0$ limit.~Since the vacuum polarization is computed strictly in the Minkowski limit, it knows nothing about de Sitter and is independent of $H$ which we expect plays a role in the infrared part of the calculation, as we'll see explicitly below.~Note also that one cannot invoke renormalization of the electric charge to simply throw away the $\ln (m_\phi^2/H^2)$ term, as advocated in~\cite{Banyeres:2018aax}, since this term is needed to ensure the proper behavior of the renormalized current in the large $m_\phi/H$ limit, as we'll see explicitly in~\sref{scalarcurrent}. 

However, as we have discussed above, taking $p^2 = 0$ is inconsistent with the presence of a constant (dynamical) electric field in de Sitter space, independently of considerations about renormalization or the Schwinger current.~As we have shown, in order to sustain a constant (dynamical) electric field in de Sitter space one requires the photon to have a tachyonic mass $-p^2 = m_A^2 = -2H ^2$.~If we evaluate the vacuum polarization diagram now with $p^2 = 2H^2$ and again in the Minkowski limit we instead obtain,
\bea\label{Eq:del3phiapp}
\delta_3^{\phi} &=& -e^2\Pi(2H^2) = 
\left(\frac{e}{12\pi}\right)^2
\Big[ -3 \ln \Big(\frac{\Lambda ^2}{H^2}\Big)
+ 3 \ln \Big(\frac{m_\phi^2}{H^2}\Big) \nonumber\\
&+& 6\Big(2 \frac{m_\phi^2}{H^2} 
+ 1\Big)^{3/2}\coth ^{-1}
\Big(\sqrt{2 \frac{m_\phi^2}{H^2} + 1}\Big)
-12 \frac{m_\phi^2}{H^2}  - 8\Big]\, , ~~~~~~~~
\eea
where we see explicitly in the second line the new terms which are now generated.~Crucially, in the limit $m_\phi / H \to 0$ there is \emph{no negative IR divergence} and $\delta_3^{\phi}$ converges to,
\bea\label{Eq:del3phiapplim}
\delta_3^{\phi} &=&
\frac{e^2}{144\pi^2}
\Big[ -3 \ln \Big(\frac{\Lambda ^2}{H^2}\Big) + \ln8 - 8\Big]
\approx 
\frac{e^2}{144\pi^2}
\Big[ -3 \ln \Big(\frac{\Lambda ^2}{H^2}\Big) - 5.9 \Big] . 
\eea
The log divergence cancels the one in the regularized current giving a renormalized current which is both UV \emph{and} IR finite in the massless electron limit.

Repeating the analysis for a charged fermion loop we again first examine the counterterm assuming a massless photon with $p^2 = 0$,
\bea\label{Eq:d3psi1}
\delta_3^{\psi} &=& -e^2\Pi(0) 
=
\left(\frac{e}{6\pi}\right)^2
\Big[ -3 \ln \Big(\frac{\Lambda ^2}{m_\psi^2}\Big)\Big]\, ,
\eea
where $m_\psi$ is the charged fermion mass and again we see the IR divergence in the $m_\psi \to 0$ limit.~With this counterterm one reproduces the results for the renormalized current from~\cite{Hayashinaka:2016qqn} which computed the current using adiabatic subtraction.~Now imposing the renormalization condition in~\eref{del3fix2} consistent with a constant electric field in de Sitter space we obtain,
\bea\label{Eq:del3psiapp}
\delta_3^{\psi} &=& -e^2\Pi(2H^2) =
\frac{e^2}{36\pi^2 }
\Big[
-3 \ln \Big(\frac{\Lambda ^2}{H^2}\Big) 
+ 3 \ln \Big(\frac{m_\psi^2}{H^2}\Big) \\
&-& 6 \Big(\frac{m_\psi^2}{H^2} - 1\Big) 
\sqrt{2\frac{m_\psi^2}{H^2} +1}\, 
\coth ^{-1}\Big({\sqrt{2\frac{m_\psi^2}{H^2} + 1}}\Big) 
+ 6 \frac{m_\psi^2}{H^2}  - 5\Big] , \nonumber
\eea
where again in the second line we see the new terms generated by the tachyonic photon mass condition.~Again this is free of IR divergences in the $m_\psi/H \to 0$ limit which gives,
\bea\label{Eq:del3psiapplim}
\delta_3^{\psi}
 &=& 
\frac{e^2}{36\pi^2 }
\Big[
-3 \ln \Big(\frac{\Lambda ^2}{H^2}\Big)
+ \ln8  - 5\Big]
\approx
\frac{e^2}{36\pi^2 }
\Big[
-3 \ln \Big(\frac{\Lambda ^2}{H^2}\Big)
- 2.9\Big]  . 
\eea
We also see that the log divergent and log 8 terms have the same coefficient as the scalar case up to a factor of 4.~As in the scalar case, the log divergence cancels the one in the regularized current giving a renormalized fermion current which is both UV \emph{and} IR finite in the massless electron limit.~While~\eref{del3phiapplim} and~\eref{del3psiapplim} are obtained in the Minkowski and large loop momentum limit, they are perfectly self consistent and free of IR divergences which have been cured, not by hand, but by imposing the tachyonic photon mass condition needed for sustaining a constant electric field in de Sitter space.~

\subsection{Renormalized Schwinger current and modified counterterms}\label{sec:current}

Until now, our previous discussion has been completely independent of considerations of the Schwinger current.~With the counterterm defined in~\eref{del3fix2} we can define the physical renormalized current.~Varying the action of \eref{renormlag} with respect to the gauge field,
\bea \label{Eq:currentdef} 
\left(1 + \delta _3 \right)\partial^{\mu }F_{\mu \nu } -  m_A^2 A_\nu = J_{\nu }  \, .
\eea 
Using the field configuration given in \eref{Edef}, we have $\partial^\mu F_{\mu\nu} = -2 a H E \delta_\nu^z$.~While we treat $A_\mu$ as classical, we are going to compute a current arising from the quantum fluctuations of charged scalar and fermion fields.~To compare such a current to the other classical terms in~\eref{currentdef} we should take its vacuum expectation value 
$\bra{0} J_\mu \ket{0} \equiv \vev{J_\mu}$ where $\ket{0}$ is the Bunch-Davies vacuum of the charged field giving for~\eref{currentdef},
\bea\label{Eq:rencurrent2}
\partial^{\mu }F_{\mu \nu } - m_A^2 A_\nu = \vev{J_\nu}_{\rm reg} + 2 aH E \delta_\nu^z \delta_3  \, .
\eea 
We have added the subscript `reg' to the expectation value of the current since, as we will see explicitly in the next sections, $\vev{J_\mu}$ is divergent and must therefore be regularized.~We expect the structure of such divergences to be the same as those of the counterterm as we will verify explicitly.~The combination on the right hand side of \eref{rencurrent2} is then finite.~We emphasize that the same scheme must be used to regularize both $\vev{J_\mu}$ and $\delta_3$.~Furthermore, an appropriate renormalization condition must be chosen to fix once and for all the finite parts of the counterterm after which we obtain an unambiguous renormalized current,
\bea \label{Eq:Jrendef}
 \langle J_z \rangle_{\rm ren} =  \langle J_z \rangle_{\rm reg} + 2 aHE \,\delta_3 \, .
\eea 
The renormalized current clearly will be UV finite, but as we argue and demonstrate below, it should be free of any IR divergences and always positive as well. 

The regularized current is obtained non-perturbatively containing all of the effects coming from de Sitter space and full dependence on the electron mass where crucially, there is no IR divergence in the small electron mass limit.~However, the counterterm in~\eref{del3phiapplim} has been obtained taking the Minkowski limit in the vacuum polarization diagram.~Thus, although there is no IR divergence when taking $p^2 = 2H^2$, we should not expect to properly capture all IR effects.~We can see this when considering the renormalized Schwinger current where, because of the missing contributions from de Sitter space, this procedure for computing the vacuum polarization diagram leads to a small (constant) negative finite contribution to the counterterm of roughly $\delta_{3\rm{fin}}^{\phi} \approx -0.004e^2$ in~\eref{del3phiapplim} for a charged scalar and $\delta_{3\rm{fin}}^{\psi} \approx -0.008e^2$ in~\eref{del3psiapplim} for a charged fermion.~As we'll see below, in the small electron mass limit, this leads to a constant, but slightly negative renormalized current implying the unphysical behavior of a current flowing opposite to the electric field.~As we explore more below, we expect that including additional corrections to the vacuum polarization from de Sitter space ensures a positive current.

In general corrections from de Sitter space cannot be included analytically, but one that can be included analytically is the curvature correction to the electron mass coming from a non-minimal coupling to gravity in the case of a scalar or the spin connection as in the case of the charged fermion.~These corrections are automatically included in the \emph{regularized} current which is computed non-perturbatively in de Sitter space directly from the Lagrangian.~However, in the counterterm they are not included in the pure Minkowski approximation of the electron propagator used above and implicitly in previous calculations in the literature.~To see how they enter in the electron propagator we can consider the equations of motion in a curved background for a free non-minimally coupled scalar,
\bea\label{Eq:phiEOM}
(\Box + m_\phi^2 + \xi R)\phi=0 \, ,
\eea
as well as for a free fermion,
\bea\label{Eq:psiEOM}
(\Box + m_\psi^2 + \frac{1}{4} R ) \psi = 0 .
\eea
The latter is obtained~\cite{Parker:2009uva} by taking the square of the Dirac operator, $\slashed D \slashed D \psi =0$ with $\slashed D = \gamma^\mu D_\mu$ and $D_\mu = \partial_\mu + B_\mu$ where $B_\mu$ is the spin connection.~One consequence of the curved background is the addition of a term proportional to $R$ in the Klein-Gordon equation.~In particular, in de Sitter space, $R=12H^2$ is a constant and such a term can be incorporated into the squared mass term.~We account for that by taking the free scalar propagator proportional to $(k^2 + m_\phi^2 + \xi R)^{-1}$ and the free fermion propagator proportional to $(k^2 + m_\phi^2 + R/4)^{-1}$.~We use these propagators in the calculation of the one-loop diagrams where the kinematics is considered to be as in a flat background.~There are additional (time dependent) corrections from de Sitter space inside the $\Box$ operator which our calculation does not account for, but to our current knowledge they cannot be computed analytically.~Nevertheless, we expect that if one were eventually to perform the calculation of the counterterm fully analytically in de Sitter space, the resulting renormalized current would still be positive in all regimes. 

With the modified propagators we repeat the calculation of the vacuum polarization imposing the tachyonic photon mass in the renormalization condition in~\eref{del3fix2} to obtain the counterterm (see~\aref{delta3S} for details),
\bea\label{Eq:delta3PV}
\delta_3^{\phi} 
&=&
\frac{e^2}{144\pi^2}
\Big[ -3 \ln \Big(\frac{\Lambda ^2}{H^2}\Big)
+ 3 \ln \Big( \frac{m_\phi^2}{H^2} + 12\xi \Big)
-12 \Big( \frac{m_\phi^2}{H^2} + 12\xi \Big)  \nn 
&+& 6\Big(2 \Big( \frac{m_\phi^2}{H^2} + 12\xi \Big) 
+ 1\Big)^{3/2}\coth ^{-1}
\Big(\sqrt{2\Big( \frac{m_\phi^2}{H^2} + 12\xi \Big)+1}\Big)-8\Big]\,,
\eea
for the scalar case, while for the fermion case we obtain (see~\aref{delta3F} for details),
\bea\label{Eq:delta3PVf}
\delta_3^{\psi} &=& 
\frac{e^2}{36\pi^2 }
\Big[
-3 \ln \Big(\frac{\Lambda ^2}{H^2}\Big)
+ 3 \ln \Big(\frac{m_\psi^2}{H^2}+3\Big) 
+ 6 \Big(\frac{m_\psi^2}{H^2} + 3\Big)  \nn
&-& 6 \Big(\Big(\frac{m_\psi^2}{H^2} + 3\Big) - 1\Big) 
\sqrt{2\Big(\frac{m_\psi^2}{H^2}+3\Big)+1}\, \coth ^{-1}
\Big({\sqrt{2\Big(\frac{m_\psi^2}{H^2} + 3\Big) + 1}}\Big) - 5\Big] \, ,
\eea
where again $\Lambda$ is the mass of the Pauli-Villars fields and
we have taken $R = 12H^2$ for the curvature in de Sitter space.~We see explicitly that the curvature corrections to the mass give additional \emph{positive} finite contributions in the counterterm.~Now in the massless limit,

\bea\label{Eq:delta3PVlim}
\delta_3^{\phi} 
&=&
\frac{e^2}{144\pi^2}
\Big[ -3 \ln\Big(\frac{\Lambda ^2}{H^2}\Big)
+ \ln 8 - 32 + 30 \sqrt{5} \cot^{-1}(\sqrt{5}) \Big] \nn
&\approx&
\frac{e^2}{144\pi^2}
\Big[ -3 \ln \Big(\frac{\Lambda ^2}{H^2}\Big)
+ 2.4 \Big]\,,
\eea
where we have taken $\xi = 1/6$ for the case of a conformally coupled scalar.~So we see explicitly the positive contributions coming from the curvature corrections to the charged scalar mass ensure a positive finite contribution\footnote{Note for the minimally coupled scalar case with $\xi = 0$ there is no curvature correction to the mass.~In this case the finite contribution from the counterterm is negative as in~\eref{del3phiapplim}, but the regularized current is positive definite with a larger magnitude than the negative finite piece of the counterterm so the renormalized current is always positive in this case as well (see solid curves in~\fref{scalarfig1}).} to the counterterm of $\delta_{3\rm{fin}}^{\phi} \approx 0.002e^2$.~As we'll see below, this ensures a positive renormalized scalar current.~For the charged fermion counterterm in the massless limit we obtain (see~\aref{delta3F} for details), 
\bea\label{Eq:delta3PVflim}
\delta_3^{\psi} &=& 
\frac{e^2}{36\pi^2 }
\Big[
-3 \ln \Big(\frac{\Lambda ^2}{H^2}\Big)
+ \ln(27) + 13
-12\sqrt{7}\coth ^{-1}(\sqrt{7}) \Big]  \\
&\approx&
\frac{e^2}{36\pi^2 }
\Big[
-3 \ln \Big(\frac{\Lambda ^2}{H^2}\Big)
+ 3.7 \Big]  , \nonumber
\eea
which leads to a positive finite contribution to the counterterm of $\delta_{3\rm{fin}}^{\psi} \approx 0.01e^2$ and again to a positive renormalized fermion current, as we'll see below.~These counterterms will lead to renormalized currents which are both UV and IR finite as well as always positive.

Before continuing with the calculation we summarize our previous discussion:  
\begin{itemize}
\item The UV divergence is encoded in the $\ln \Lambda^2$ term.~If we were able to perform the exact calculation in de Sitter, we would find the same coefficient for this term.~In other words, the Minkowski calculation captures correctly the UV divergence structure of de Sitter.~This is simply the statement that for UV modes the curvature is irrelevant.~The $\ln \Lambda^2$ term will cancel exactly against the analogous term calculated explicitly in the current $\vev{J_z}$ which is regularized with the same scheme. 
\item If we imposed the (unphysical) renormalization condition $\Pi(p^2 = 0) = 0$ for a massless photon in place of~\eref{del3fix2} and dropped the curvature corrections to the squared masses in the propagators we would obtain the counterterms
$\delta_3^\phi = - e^2 / (48 \pi^2) \ln(\Lambda^2 / m_\phi^2)$ and  
$\delta_3^\psi = - e^2 / (12 \pi^2) \ln(\Lambda^2 / m_\psi^2)$.~These would then lead to renormalized currents which would match exactly those found in previous literature~\cite{Kobayashi:2014zza,Hayashinaka:2016qqn,Banyeres:2018aax} and reproduce the same puzzling negative IR divergences. 
This points clearly to the issue of those derivations. 
\item Taking the limit $m_{\phi, \psi} \to 0$ and $R\to 0$, that is dropping the squared mass and its curvature correction, we still have counterterms free of IR divergences.~What cures such divergences, which were erroneously found in previous literature~\cite{Hayashinaka:2016qqn}, is the renormalization condition (\eref{del3fix2}) imposed at $p^2 = -m_A^2$ rather than at $p^2 = 0$.  
\item As we examine explicitly below, if we did not include the curvature correction terms in the propagators, the finite term in $\delta_3$ is negative and leads to a renormalized current which would become negative in some IR regions with $m_{\phi,\psi} / H \ll 1$.~This is a justification a posteriori for keeping those terms in the propagators, but it is clear that the full de Sitter calculation
of $\delta_3$ would take them into account.~As we'll show below, including them in the Minkowski calculation leads to a renormalized current which is always well behaved in the IR and always positive, as expected. 
\item Given our procedure, the renormalized gauge coupling $e$, which will enter in many results in what follows, is the electric charge of the charged particles that an observer would measure at around the Hubble scale, $|p| = |m_A| = \sqrt{2} H$.~The gauge coupling at a different energy scale is readily obtained by running with the beta function which is the same as in flat spacetime. 
\end{itemize}

\section{Scalar Schwinger current in de Sitter}\label{sec:scalarcurrent}

In this section we summarize the calculation of the charged scalar Schwinger current in de Sitter space.~We refer the reader to~\cite{Kobayashi:2014zza,Banyeres:2018aax} for more details of the derivation.~As we comment more below, there are important differences with the results obtained in this study and those in the literature due to different renormalization procedures.

Our starting point is the action for scalar QED in de Sitter space,
\bea\label{Eq:actionSQED} 
S &=& \int d^4 x \sqrt{-g}
\Big[-\frac{1}{4}(1+\delta_3)F^{\mu\nu}F_{\mu\nu}-\frac{1}{2} m_A^2 A_\mu A^\mu\nn
&-& g^{\mu \nu}\left(\partial_\mu-i e A_\mu\right) \phi^*\left(\partial_\nu+i e A_\nu\right) \phi-(m_\phi^2+\xi R) \phi^* \phi  \Big] .
\eea
We assume a constant and uniform external electric field $E$ along the $z$-direction direction with the gauge field given by \eref{Edef} and $m_A^2 = -2 H^2$.~After redefining the scalar field to $\phi = q / a$ one obtains the equation of motion for the mode functions $q_k$~\cite{Kobayashi:2014zza,Banyeres:2018aax},
\bea
q_k^{\prime \prime}+\omega_{\boldsymbol{k}}^2 q_{\boldsymbol{k}}=0,
\label{Eq:EOMscalarmodes}
\eea
where primes indicate derivatives with respect to $\tau$.~The mode frequency $\omega_k$ is,
\bea\label{Eq:omegak}
	\omega_k^2 &=&
	\left(k_z+e A_z\right)^2		
	+ k_x^2 + k_y^2 + a^2 (m_\phi^2+\xi R) -\frac{a^{\prime \prime}}{a} \\
	&=& k^2 + 2 (aH) \lambda r k 
	+ (aH)^2 \left( 1/4 - \mu^2 \right) \nonumber,
\eea
where we defined $k\equiv \left(k_x^2+k_y^2+k_z^2\right)^{1/2}$ as well as the dimensionless ratios,
\bea
\lambda=\frac{e E}{H^2} , 
\quad \mbphi = \frac{m_\phi}{H} ,
\quad \mu^2=\frac{9}{4} - (\mbphi^2 + 12\xi) - \lambda^2 ,
\quad r=\frac{k_z}{k} \, ,
\eea
and used $R=12\, H^2$ for the Ricci scalar in de Sitter space.~We can see explicitly in~\eref{omegak} that even in the case with $m_\phi = 0$ and $\xi = 1/6$, which gives $\mu^2 = 1/4 - \lambda^2$, there is still a source of conformal symmetry breaking through $\lambda$, i.\,e.\,the constant electric field.~Thus conformally coupled (charged) scalars, which are not gravitationally
produced, can still be produced via the Schwinger mechanism.~The exact solution for the equation of motion in~\eref{EOMscalarmodes} is known in terms of the Whittaker function,
\bea
q_k=\frac{\mathrm{e}^{-\pi \lambda r / 2}}{\sqrt{2 k}} W_{i \lambda r, \mu}(2 i k \tau)\,,
\eea
where $W_{i\lambda r,\mu}(2ik\tau)$ is the positive frequency solution in the asymptotic past $(-k\tau\rightarrow \infty )$. 

Varying the action with respect to the gauge field leads to~\eref{currentdef} with the conserved current for a charged scalar given by,
\bea
J^\phi_\mu = \frac{i e}{2}\left\{\phi^{\dagger}\left(\partial_\mu+i e A_\mu\right) \phi-\phi\left(\partial_\mu-i e A_\mu\right) \phi^{\dagger}\right\}+\text { h.c. }\,.
\eea
This is an operator defined in terms of the quantized scalar field $\phi$ and we are interested in its expectation value.~With a constant electric field in the $z$-direction, the expectation value of the current will vanish in all components except in the $z$-direction giving,
\bea
\vev{ J^\phi_z} \equiv
\bra{0} J^\phi_z \ket{0} =\frac{2 e}{a^2} \int \frac{d^3 k}{(2 \pi)^3}\left(k_z+e A_z\right)\left|q_k\right|^2 .\label{Eq:currentVEV}
\eea
The calculation of the integral is difficult and contains quadratic and logarithmic divergences, but remarkably can be carried out analytically~\cite{Kobayashi:2014zza}.~To examine these divergences it is convenient to introduce temporarily a momentum cut off $\zeta$ which leads to,
\bea
\left\langle J^\phi_z\right\rangle 
= a H \frac{e^2 E}{4 \pi^2} \lim _{\zeta \rightarrow \infty}\left[\frac{2}{3}\left(\frac{\zeta}{a H}\right)^2+\frac{1}{3} \ln \frac{2 \zeta}{a H}-\frac{25}{36}+\frac{\mu^2}{3}+\frac{\lambda^2}{15}+F_\phi(\lambda, \mu)\right],
\label{Eq:scalardivcurrentVEV}
\eea
where we have defined the function,
{\small
\bea\label{Eq:F_phi}
F_\phi(\lambda, \mu) &\equiv&  
\frac{45+4 \pi^2\left(-2+3 \lambda^2+2 \mu^2\right)}{12 \pi^3} \frac{\mu \cosh (2 \pi \lambda)}{\lambda^2 \sin (2 \pi \mu)}
- \frac{45+8 \pi^2\left(-1+9 \lambda^2+\mu^2\right)}{24 \pi^4} \frac{\mu \sinh (2 \pi \lambda)}{\lambda^3 \sin (2 \pi \mu)} \nonumber\\
&+& 
\operatorname{Re}\Big[\int_{-1}^1 d r \frac{i}{16 \sin (2 \pi \mu)}
\left(-1+4 \mu^2+\left(7+12 \lambda^2-12 \mu^2\right) r^2-20 \lambda^2 r^4\right) \\
&\times&
\left(\left(\mathrm{e}^{-2 \pi r \lambda}+\mathrm{e}^{2 \pi i \mu}\right) \psi\left(\frac{1}{2}+\mu-i r \lambda\right)
- \left(\mathrm{e}^{-2 \pi r \lambda}+\mathrm{e}^{-2 \pi i \mu}\right) 
\psi\left(\frac{1}{2}-\mu-i r \lambda\right)\right)\Big] .\nonumber
\eea}

Following~\cite{Banyeres:2018aax}, we regularize the current with the manifestly gauge invariant Pauli-Villars method.~We introduce 3 extra heavy fields which satisfy the sum rules~\cite{Banyeres:2018aax},
\bea
\sum_{i=0}^3 (-1)^i = 0 \quad \text { and } \quad 
\sum_{i=0}^3 (-1)^i m_i^2=0,\label{Eq:PVcond1}
\eea
where $i = 0$ represents the original charged field $\phi$ which has mass $m_0^2 = m_\phi^2 + 12 \xi H^2$ while the auxilary fields have masses $m_2^2 = 4\Lambda^2-m_0^2$ and $m_1^2=m_3^2=2\Lambda^2$.~Here $\Lambda$ plays the role of the regulator which will be sent to infinity at the end of the calculation.~The first sum enforces two of the extra heavy fields to have the wrong kinetic sign.~Using \eref{scalardivcurrentVEV} for each field and summing the contributions we obtain the regularized current.
\begin{equation}
	\left\langle J^\phi_z\right\rangle_{\mathrm{reg}}
	= \sum_{i=0}^3 (-1)^i  \left\langle J_z\right\rangle_{i}
	= a H \frac{e^2 E}{4 \pi^2}  
	\left[\frac{1}{6} \ln \Big(\frac{\Lambda^2}{H^2}\Big)
	-\frac{2 \lambda^2}{15}+F_\phi(\lambda, \mu)\right] \, ,
	\label{Eq:PVcond2}
\end{equation}
Using the counterterm of~\eref{delta3PV}, derived by fixing the renormalization condition of~\eref{del3fix2}, we obtain the physical renormalized Schwinger current,
\begin{flalign}\label{Eq:Jphiren}
	\langle J_z^\phi \rangle_{\rm ren} &
	= \langle J^\phi_z \rangle_{\mathrm{reg}} 
+ (2aH E)\delta^\phi_3\nonumber \\
	& = a e H^3 \frac{\lambda}{4 \pi^2}  
	\left[
	- \frac{2 \lambda^2}{15}
	+ F_\phi(\lambda, \mu)
	+ \frac{1}{6}\ln{ ( \mbphi^2 + 12\xi ) }
	- \frac{2}{3}( \mbphi^2 + 12\xi ) 
	\right.\\ 
	& \left.+\frac{1}{3}\left(2 ( \mbphi^2 + 12\xi ) + 1\right)^{3/2}
	\coth ^{-1}\left(\sqrt{2 ( \mbphi^2 + 12\xi )  +1}\right) - \frac{4}{9} \right] ,\nonumber
\end{flalign}
which is \emph{finite and positive} for all parameters (see~\fref{scalarfig1}).

As anticipated, the $\ln \Lambda^2$ from $\vev{J_z}_{\rm reg}$ cancels exactly against the same term in the counterterm.~If we set $\xi = 0$, the first three terms match exactly the results found in previous calculations using different regularization and renormalization procedures:\,in~\cite{Kobayashi:2014zza} which utilized adiabatic subtraction to renormalize the current, in~\cite{Hayashinaka:2016dnt} which utilized the point-splitting scheme, and in~\cite{Banyeres:2018aax} which used RG evolution taking the renormalization scale to be the electron mass $m_\phi$.~These first three terms can also be obtained from our calculation by setting $p^2 = 0$ (massless photon) in the vacuum polarization diagram which leads to $\delta^\phi_3 = - \frac{e^2}{48\pi^2} \ln\frac{\Lambda^2}{m_\phi^2}$.~So we see these various methods of regularization and renormalization, which all give the same result for the renormalized current, are equivalent to computing the vacuum polarization diagram and imposing the renormalization condition in~\eref{del3fix2}, but setting $p^2 = -m_A^2 = 0$.~However, if we keep only the first three terms (with $\xi = 0$) in~\eref{Jphiren}, then we see there is a leftover (negative) logarithmic divergence in the $m_\phi \to 0$ limit as there is no longer a $\coth^{-1}\sqrt{\mbphi^2 + 1}$ term to cancel it.~So we can trace the peculiar negative IR divergence in the renormalized scalar current found in previous literature~\cite{Kobayashi:2014zza,Hayashinaka:2016dnt,Banyeres:2018aax}\,\footnote{The authors of~\cite{Banyeres:2018aax} argue that one can run the electric charge from the electron mass to the Hubble scale in order to remove the negative IR divergent log term and obtain a positive current when the mass is small.~This is akin to simply dropping the log divergent term in~\eref{PVcond2}.~However, when the mass is large this log term must be kept to obtain the correct behavior in the large mass limit so it is not clear to us that this is an unambiguous procedure.~Furthermore, since it is the renormalized electric charge which appears in the current, it is unclear why further renormalization should be needed to remove IR divergences.} to unphysical renormalization conditions that are inconsistent with a constant classical electric field in de Sitter space which, as shown in~\sref{efield}, requires a photon with a tachyonic mass $p^2 = -m_A^2 = 2H^2$.

\subsection{Behavior of the scalar current}\label{sec:scalarlimit}
 
With the result for the induced comoving Schwinger current in hand we now examine its behavior and study different limits.~The physical current is defined as $J = \langle J_z^\phi \rangle_{\rm ren} / a$.~It is convenient to define a dimensionless current which we will examine in what follows,
\bea\label{Eq:dimlessJphi}
\mathcal{J}_\phi \equiv  \frac{\langle J_z^\phi \rangle_{\rm ren}}{a e H^3} ,
\eea
and which depends only on $\lambda$, $\mbphi$, and $\xi$.~In~\fref{scalarfig1} we show $\mathcal{J}_\phi$ for a minimally coupled scalar with $\xi=0$ (solid) and a conformally coupled scalar with $\xi = 1/6$ (dashed) as a function of $\lambda$ (top) for various fixed values of $\mbphi$ and as a function of $\mbphi$ (bottom) for various fixed values of $\lambda$.~We can then examine the behavior in various limits. 
\begin{figure}[ht]
\centering
\includegraphics[totalheight=9.5cm]{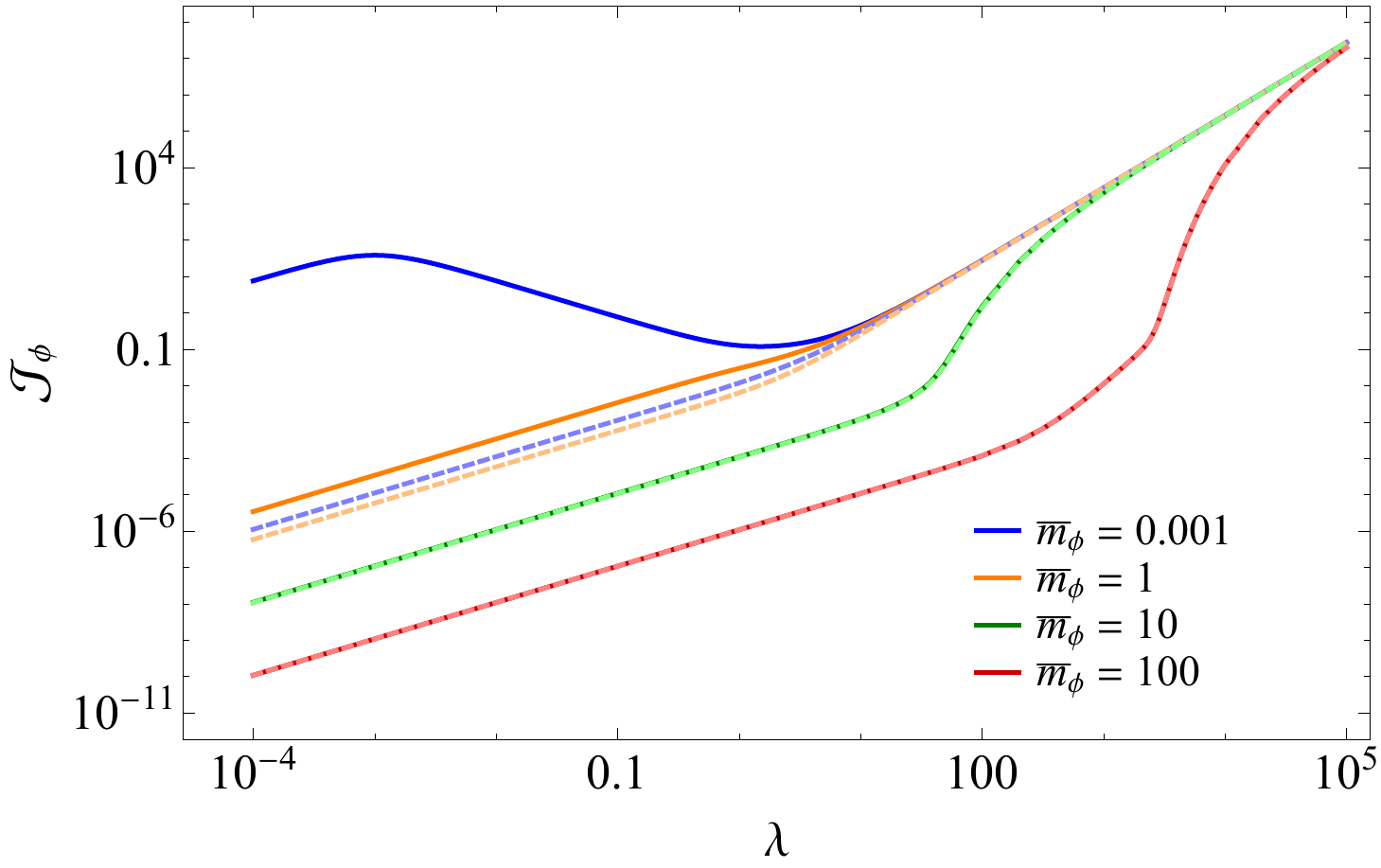} 
\includegraphics[totalheight=9.5cm]{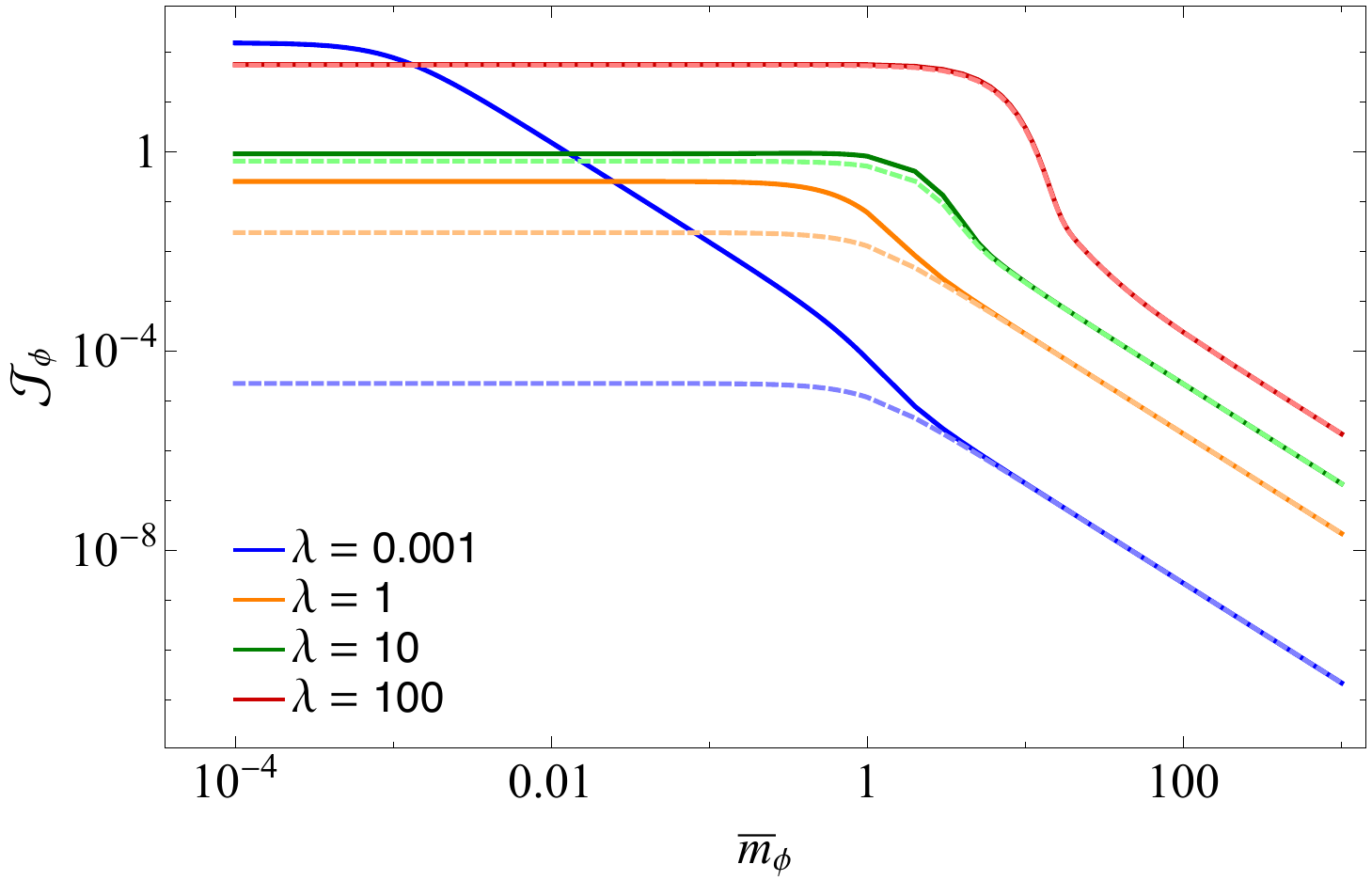} 
\caption{{\bf Top:\,}Induced (dimensionless) Schwinger current defined in~\eref{dimlessJphi} for a minimally coupled scalar with $\xi=0$ (solid) and a conformally coupled scalar with $\xi = 1/6$ (dashed) as a function of $\lambda \equiv eE/H^2$ for various values of fixed $\mbphi \equiv m_\phi/H$ as labeled in the figure.\,{\bf Bottom:\,}Same as top but now as a function of $\mbphi$ for various fixed values of $\lambda$.}
\label{fig:scalarfig1}	
\end{figure}

\subsubsection{Strong force limit: $|eE| \gg H^2$}

In the limit $\lambda\gg 1$ and $\lambda \gg \mbphi^2 + 12\xi$ the function $F_\phi(\lambda,\mu)$ given in \eqref{Eq:F_phi} approaches,
\begin{equation}
F_\phi(\lambda,\mu) \to \frac{2}{15}\lambda^2 + \frac{\lambda}{3\pi} e^{-\pi \frac{-9/4 + \mbphi^2 + 12\xi}{\lambda}} \, .
\end{equation} 
The term proportional to $\lambda^2$ here cancels exactly against the first term in \eref{Jphiren}.~The remaining terms in~\eref{Jphiren} which come from the counterterm give subdominant contributions and can be neglected.~The dimensionless current then goes as,
\bea \label{Eq:Jphistrong}
\mathcal{J}_\phi 
\simeq  \frac{\lambda^2}{12 \pi^3} e^{-\pi \frac{-9/4 + \mbphi^2 + 12\xi}{\lambda}} \, .
\eea
It is instructive to compare it to the result derived with a different method to calculate the current in this limit.~Under the assumption $\lambda^2 + \mbphi^2+12\xi \gg 1$, there exists an adiabatic vacuum for the charged scalar in the asymptotic future and one can calculate the pair production rate
by computing Bogoliubov coefficients as was done in~\cite{Kobayashi:2014zza}.~Repeating their analysis, one finds the number of pairs produced per unit physical four-volume $\Gamma = H^4/(2\pi)^3 \lambda^2 \exp(-\pi(\mbphi^2+12\xi - 9/4)/\lambda^2)$.~From this constant rate we have the number density of pairs $n = \Gamma / (3H)$ and we can estimate the physical current as $J = 2 |e| n v$.~Here the factor 2 accounts for the fact that particles are produced in pairs and $v$ is their velocity which we can take to be
$v \sim 1$ in the strong electric field.~Then the current is in perfect agreement with
$J = \langle J_z^\phi \rangle_{\rm ren} / a = e H^3 {\cal J}_\phi$ where ${\cal J}_\phi$ is given by~\eref{Jphistrong}.

At large $\lambda$ the exponential in~\eref{Jphistrong} is of order one, so $\mathcal{J}_\phi $ approaches $\lambda^2/(12\pi^3)$ for any value
of the scalar electron mass as $\lambda$ grows.~We can see this explicitly in~\fref{scalarfig1}.~In the limit $\lambda \gg  \mbphi^2 \gg (1, 12\xi)$, the terms $-9/4$ and $12\xi$ in the exponent can be neglected and one obtains the physical current,
\bea
\frac{\langle J_z^\phi \rangle_{\rm ren}}{a} = \frac{1}{12\pi^3} \frac{|e|^3 E^2}{H} e^{-\pi \frac{m_\phi^2}{|e E|}} \, .
\eea 
Replacing $1/H \to t - t_0$, thinking of $t_0$ as the time when the electric field is turned on, we recover the expression for the Schwinger current in flat spacetime as found in~\cite{Kobayashi:2014zza,Banyeres:2018aax}.~This expression diverges for $H\to 0$.~As noted in~\cite{Kobayashi:2014zza}, this is due to the fact that in the absence of Hubble dilution, the current would diverge in Minkowski space if the electric field existed in the infinite past, in contrast to the situation in de Sitter space, where the electric field can be assumed to always exist and the current is finite.

\subsubsection{Weak force limit: $|eE| \ll H^2$}

In the weak field limit $\lambda \ll 1$ we find,
\begin{equation} \label{Eq:smallambda}
\begin{split}
\mathcal{J}_\phi \simeq 
\frac{\lambda}{72 \pi ^2} & \left[ 3 \log \left(\mbphi^2+12 \xi \right)
 +6 \left(2
   \mbphi^2+24 \xi +1\right)^{3/2} \coth ^{-1}\left(\sqrt{2 \mbphi^2+24
   \xi +1}\right)
    \right. \\
   & \left. -12 \mbphi^2-144 \xi
   -8 - \frac{2
   \pi  \sqrt{-4 \mbphi^2-48 \xi +9} \left(4 \mbphi^2+48 \xi -5\right)}{
   \sin \left(\pi  \sqrt{-4 \mbphi^2-48 \xi +9}\right)} \right. \\
   & \left. 
   -3 \psi
   \left(\frac{1}{2}-\frac{1}{2} \sqrt{-4 \mbphi^2-48 \xi
   +9}\right) - 3 \psi \left(\frac{1}{2}+\frac{1}{2} \sqrt{-4 \mbphi^2-48 \xi
   +9}\right)\right] \, .
\end{split}
\end{equation}
The current is proportional to $\lambda$ and therefore linear in the electric field.~Let us consider next taking $\mbphi = 0$; the quantity in the squared parentheses in~\eref{smallambda} is positive for any $\xi > 0$.~For $\xi = 1/6$, the conformal case, we obtain,
\bea\label{Eq:JScon}
\mathcal{J}_\phi \simeq 
\frac{\lambda}{12 \pi^2}
\left(
(\gamma_E + \frac{1}{3}) 
+ (\frac{\ln 2}{2} - 4
+ 5 \sqrt{5} \coth^{-1}(\sqrt{5})
- \frac{4}{3}) \right) \, ,
\eea
where $\gamma_E \approx 0.577$ is Euler's constant.~The two terms in the first parenthesis come from the regularized current while the terms in the second parenthesis come from the counterterm.~The first three terms in the second parenthesis arise due to the curvature corrections to the electron mass in the propagator.~So we see explicitly the curvature corrections to the conformally coupled scalar mass entering in $\delta^\phi_3$ are crucial (otherwise the $\log$ and $\coth^{-1}$ terms cancel as $\mbphi\to0$) for compensating the negative term and ensuring a positive renormalized current.~We also see that in the massless limit the current for a conformally coupled scalar converges to a constant that is fixed by the physical renormalization in~\eref{del3fix2} and numerically equal to $\mathcal{J}_\phi \approx 0.011\lambda$.~As we'll see in the next section, for a conformally coupled scalar the physical current in~\eref{Jphiren} has the same behaviour as the charged fermion current (see~\fref{fermionfig1}), which gives us further confidence in our renormalization procedure.

It is instructive to also consider the case with $\xi = 0$ and $\mbphi^2 + \lambda^2 \ll 1$.~In this limit the main contribution to the current comes from $F_\phi(\lambda,\mu)$ and one finds,
\bea \label{Eq:JscalHypercond}
\mathcal{J}_\phi 
\simeq \frac{3}{4\pi^2}\frac{\lambda}{\lambda^2 +  \mbphi^2} \, .
\eea
When $\mbphi^2 \ll \lambda^2$ the current goes as $\lambda^{-1}$ so it increases as the electric field becomes weaker.~This surprising behavior, first noted in $1+1$ dimensions~\cite{Frob:2014zka} then in $3+1$ dimensions~\cite{Kobayashi:2014zza,Banyeres:2018aax}, is known 
as infrared hyperconductivity and is a feature found in de Sitter space, but absent
in flat space.~In~\cite{Banyeres:2018aax} a nice explanation for this phenomenon was provided which we summarize here.~A light scalar field experiences gravitational particle production in a de Sitter background, leading to the expectation value $\langle \phi^* \phi \rangle \simeq 3 H^4 / (4 \pi^2 m_\phi^2)$.~In the presence of a weak electric force the squared mass gets effectively replaced~\cite{Frob:2014zka} by $m_\phi^2 \to m_\phi^2 + e^2 E^2 / H^2$.~The expectation value of the scalar field then gives mass to the vector field $m^2_A = e^2 \langle \phi^* \phi \rangle \simeq 3 e^2 H^2 / (4\pi^2 (\mbphi^2 + \lambda^2))$ which turns the medium into a superconductor.~Note that since $\mbphi^2 + \lambda^2 \ll 1$ the vector induced squared mass is positive and much larger than $H^2$ so this contribution overcomes the negative tachyonic mass introduced earlier.~Writing the equation of motion for a massive gauge field, $\partial_\mu F^{\nu\mu} = -m_A^2 A^\nu \equiv J^\nu$ and plugging in the positive induced $m_A^2$, as well as the field configuration~\eref{Edef}, one obtains a current in perfect agreement with~\eref{JscalHypercond}.

\vspace{-0.12cm}
\subsubsection{Large mass limit}

We now consider the case in which the charged scalar particles are very heavy.~This limit was already largely discussed in~\cite{Banyeres:2018aax}, where it was pointed out that the resulting current
matches the one obtained from the Euler-Heisenberg effective action.~Although our final results are similar, our derivation has some differences compared to that of~\cite{Banyeres:2018aax}.~First, we separate the contribution of the counterterm which should not be included in the comparison, as we explain below;
second, we consistently keep track from the very beginning of the terms proportional to the $\xi$ coupling of the scalar field to the Ricci scalar.~The latter point results in several terms with powers of $\xi$ which were not accounted for in~\cite{Banyeres:2018aax}. 

Starting from~\eref{Jphiren} and taking the limit $\mbphi^2 \gg (\lambda, 12 \xi)$ gives
\bea \label{Eq:JreglarM1}
\left\langle J^\phi_z\right\rangle_{\mathrm{ren}}  
&=&  a e H^3 \frac{\lambda}{4\pi^2}  
\Big[ F_\phi(\lambda,\mu)\vert_{\mbphi \gg 1} - \frac{2\lambda^2}{15} + \frac{1}{3} \ln \mbphi  \nonumber \\ 
&+& \frac{1}{\mbphi^2} \left( \frac{1}{30} + 2\xi \right)
- \frac{1}{\mbphi^4} \left(\frac{1}{210} + \frac{2}{5} \xi + 12 \xi^2 \right)  \nonumber\\
&+& \frac{1}{\mbphi^6} \left( \frac{1}{945} + \frac{4}{35} \xi + \frac{24}{5} \xi^2 + 96\xi^3  \right) \\
&-& \frac{1}{\mbphi^8} \left( \frac{1}{3465} + \frac{4}{105} \xi + \frac{72}{35} \xi^2 + \frac{288}{5} \xi^3 + 864 \xi^4  \right) + {\cal O}\left(\frac{1}{\mbphi^{10}} \right)\Big] ,\nonumber
\eea
where the integral function is given by,
\bea\label{Eq:FlarM}
&&F_\phi(\lambda,\mu)\vert_{\mbphi \gg 1} =
\frac{2\lambda^2}{15} - \frac{1}{3} \ln \mbphi \nonumber\\
&&~+ \frac{1}{\mbphi^2} \left( \frac{7}{18} - 2\xi \right) + \frac{1}{\mbphi^4} \left( \frac{41}{90} + \frac{7}{180} \lambda^2 -\frac{14}{3} \xi +12\xi^2  \right) \nonumber\\
&&~+ \frac{1}{\mbphi^6} \left( \frac{676}{945} +\frac{19}{90} \lambda^2 - \frac{164}{15} \xi - \frac{14}{15} \lambda^2 \xi + 56 \xi^2 - 96 \xi^3  \right)  \\
&&~+ \frac{1}{\mbphi^8} \left( \frac{401}{315} + \frac{2809}{3150} \lambda^2 + \frac{31}{840} \lambda^4 - \frac{2704}{105} \xi - \frac{38}{5} \lambda^2 \xi + \frac{984}{5} \xi^2 + \frac{84}{5} \lambda^2 \xi ^2 - 672 \xi^3  + 864 \xi^4 \right) \nonumber  \\
&&~+ {\cal O}\left(\frac{1}{\mbphi^{10}} \right) \, .\nonumber
\eea
In this limit the contribution to $F_\phi$ comes from the integral in the second and third lines of \eref{F_phi}.~In \eref{JreglarM1} the terms $-2\lambda^2/(15) + 1/3 \ln \mbphi$ in the first line cancel out while the remaining terms in lines 2, 3, and 4, suppressed by negative even powers of $\mbphi$, trace back to the contribution proportional to the counterterm $\delta_3$ in \eref{Jrendef}.

It is interesting to consider how to interpret the current in this large mass limit.~Starting with the action of \eref{actionSQED} and integrating out the heavy charged scalar field we obtain,
\begin{equation} \label{Eq:SscalarEH}
S=\int d^4 x \sqrt{-g}  \left[-\frac{1}{4}(1+\delta_3)F^{\mu\nu}F_{\mu\nu}-\frac{1}{2} m_A^2 A_\mu A^\mu + 
{\cal L}_{\rm flat}^{\rm eff} + {\cal L}_{\rm curv}^{\rm eff} \right] \, ,
\end{equation}
where we have the flat-space Euler-Heisenberg Lagrangian,
\begin{equation} \label{Eq:EHflat}
\begin{split}
{\cal L}_{\rm flat}^{\rm eff} = & \frac{e^4}{23040 \pi^2 m_\phi^4} \left[ 7 \left( F_{\mu\nu} F^{\mu\nu} \right)^2 + \left( F_{\mu\nu} \tilde F^{\mu\nu} \right)^2 \right] \\
&  -  \frac{e^6}{80640 \pi^2 m_\phi^8} F_{\mu\nu} F^{\mu\nu}
\left[ \frac{31}{4} \left( F_{\mu\nu} F^{\mu\nu} \right)^2 + \frac{77}{16} \left( F_{\mu\nu} \tilde F^{\mu\nu} \right)^2 \right] 
+ {\cal O}\left( \frac{1}{m_\phi^{12}} \right)  \, ,
\end{split}
\end{equation}
as well as the Euler-Heisenberg Lagrangian in curved-space,
\begin{equation} \label{Eq:EHcurv}
\begin{split}
{\cal L}_{\rm curv}^{\rm eff} =  \frac{e^2}{16 \pi^2 m_\phi^2} & \left[ \frac{1}{12} \left(\xi - \frac{1}{6} \right) R F_{\mu\nu} F^{\mu\nu} - \frac{1}{90} R_{\mu\nu} F^{\mu \alpha} {F^\nu}_\alpha \right. \\
& \left. -\frac{1}{180} R_{\mu\nu\alpha\beta} F^{\mu\nu} F^{\alpha \beta} + \frac{1}{60} \partial^\alpha F_{\alpha\mu} \partial_\beta F^{\beta\mu}  \right] 
+ {\cal O}\left( \frac{1}{m_\phi^{4}} \right)  \, ,
\end{split}
\end{equation}
which was derived in~\cite{Bastianelli:2008cu} for a generic curved background.~The only degrees of freedom now are those of the gauge field $A_\mu$ which we are still treating as classical so we still only need one counterterm in $\delta_3$.~Varying the action of \eref{SscalarEH} 
with respect to $A_\mu$ we obtain,
\begin{equation}\label{Eq:EOMscaleff}
(1+\delta_3) \partial^\mu F_{\mu\nu} - m_A^2 A_\nu = J_{{\rm flat},\nu}^{\rm eff} + J_{{\rm curv},\nu}^{\rm eff} \, ,
\end{equation}
where the effective currents, after setting $A_\mu = E \delta_\mu^z / (H^2 \tau)$, are given by,
\bea
J_{{\rm flat},\mu}^{\rm eff} &=& a e H^3 \frac{\lambda}{4\pi^2} \left( \frac{7}{180}\frac{\lambda^2}{\mbphi^4} + \frac{31}{840} \frac{\lambda^4}{\mbphi^8}  \right) + \dots ,\nonumber\\  
J_{{\rm curv},\mu}^{\rm eff} &=& a e H^3 \frac{\lambda}{4\pi^2} \frac{1}{\mbphi^2} \left( \frac{7}{18} - 2\xi \right) + \dots \, .
\eea
Note that $J_{{\rm flat},\mu}^{\rm eff}$ is proportional to $H$ and $J_{{\rm curv},\mu}^{\rm eff}$ to $H^3$, so both vanish when $H \to 0$, implying that in the case under consideration with a constant electric field they only play a role in a curved background.~We want to compare these effective currents against~\eref{JreglarM1} which we derived treating the charged scalar fields as dynamical and quantized degrees of freedom, taking the large mass limit only at
the end of the calculation.~Looking at~\eref{EOMscaleff} it is clear that we should compare $J_{{\rm flat},\mu}^{\rm eff}$ and $J_{{\rm curv},\mu}^{\rm eff}$ against the terms in our regularized current, $\left\langle J^\phi_z\right\rangle_{\mathrm{reg}}$, {\em i.e.} only the terms in the first line of \eref{JreglarM1}.~The remaining terms in \eref{JreglarM1} are proportional to $\delta_3$ and are already accounted for by the
term $\delta_3 \partial^\nu F_{\mu\nu}$ in \eref{EOMscaleff}.~So for the comparison we have to check $aeH^3 \lambda F_\phi(\lambda,\mu)\vert_{\mbphi \gg 1} / (4\pi^2)$, with $F_\phi(\lambda,\mu)\vert_{\mbphi \gg 1}$ given in \eref{FlarM}, against the effective currents.~The term which goes as $\mbphi^{-2}$ matches exactly the one in $J_{{\rm curv},\mu}^{\rm eff}$.~It is also
easy to find the terms proportional to $\mbphi^{-4}$ and $\mbphi^{-8}$ in \eref{FlarM} that match those in
$J_{{\rm flat},\mu}^{\rm eff}$.

We have found that the result for the regularized current, once we take the large mass limit for the charged field, reproduces what we would obtain from
the Euler-Heisenberg Lagrangian.~Recall that computing $\langle J_\mu \rangle$ means computing the expectation value of an operator which in turn is defined in terms of creation and annihilation operators describing the quantum fluctuations
of the scalar field.~In the large mass limit these quantum fluctuations are too heavy to become real pairs of charged particles.~However, they still contribute to the expectation value of the current operator and the result matches what one would obtain by first integrating out the heavy field and including the effects
of its fluctuations in the effective Euler-Heisenberg Lagrangian.~While the flat part of this effective Lagrangian has been long known to all 
orders, the curved part is difficult to derive and to our knowledge is known only
to leading $\mbphi^{-2}$ order~\cite{Bastianelli:2008cu}.~Given that our method reproduces exactly this lowest order we speculate that it also reproduces correctly
all the higher orders in $\mbphi^{-2n}$ (with $n$ a positive integer), but leave a detailed investigation of this point to future work.

Besides the contributions suppressed by powers of $m_\phi$ just discussed, in the large mass limit there are also contributions that are exponentially suppressed, thus negligibly smaller.~However, these contributions could be interesting as perhaps they describe the production and motion of real particles, as opposed to the expectation value of the current operator from fluctuations of virtual particles.~The leading term of the exponentially suppressed contributions comes from the first line of \eref{F_phi}, and gives,
\begin{equation}\label{Eq:JlarMexp}
\left\langle J^\phi_z\right\rangle_{\mathrm{reg}}^{\rm exp} = a e H^3 \frac{\lambda}{4\pi^2} \left( - \frac{1}{3\pi^2} \frac{e^{4\pi\lambda}(2\pi \lambda -1) + 2\pi \lambda +1}{\lambda^3} \mbphi^3 e^{-2\pi \mbphi} \right) \, .
\end{equation}
This is negative for any $\lambda > 0$ and for $\lambda \ll1$ becomes,
\begin{equation}\label{Eq:JlarMexpsmallam}
\left\langle J^\phi_z\right\rangle_{\mathrm{reg}}^{\rm exp} = a e H^3 \lambda \left( - \frac{4}{9\pi} \mbphi^3 e^{-2\pi \mbphi} \right) +{\cal O}(\lambda^2) \, ,
\end{equation}
which is in agreement with~\cite{Banyeres:2018aax}, but whose interpretation is puzzling.~The form of the exponent $e^{-m_\phi / H}$ is that of a Boltzmann suppression and seems to suggest that this contribution describes real heavy pairs produced gravitationally.~However, because of the negative sign these pairs would then generate a current opposite to the electric field which is strange.~Note in~\eref{JlarMexp} we could increase $\lambda$  (within the limits $ \bar{m}_\phi\gg \lambda\gg 1$), that is the electric field, which would make the current more and more negative.~Moreover, one would expect the gravitational production to give a number density $n \sim H^3 \mbphi^{3/2} e^{-\mbphi}$ of pairs with non relativistic velocities $v \sim \lambda / \mbphi \ll 1$.~Then the current $J \propto n v \sim H^3 \lambda \mbphi^{1/2} e^{-\mbphi}$ would be proportional to a different
power of $\mbphi$ compared to $\left\langle J^\phi_z\right\rangle_{\mathrm{reg}}^{\rm exp}$.~Hence the gravitational production interpretation is difficult to reconcile.~This puzzling behavior was already pointed out and discussed in the $1+1$ dimensional case in~\cite{Frob:2014zka}.

In this limit, the Bogoliubov analysis performed in~\cite{Kobayashi:2014zza} and mentioned above in the strong force limit should hold in principle.~One could again compute the pair production rate of real pairs obtaining the number density and then the physical current $ J \simeq 2 e n v$.~Going through this exercise one obtains,
\begin{equation}
\left\langle J^\phi_z\right\rangle_{\rm Bog} = a e H^3 \lambda \frac{1}{3\pi^2} \mbphi^2 e^{-2\pi \mbphi} + {\cal O}(\lambda^3) \, .
\end{equation} 
This is again different from~\eref{JlarMexpsmallam}: it has a positive sign and is proportional to $\mbphi^2$ as opposed to $\mbphi^3$ and thus is smaller in magnitude.~There is yet another way of computing this current via a semi-classical calculation for which we refer the reader to section 5 of~\cite{Banyeres:2018aax} which was inspired by~\cite{Frob:2014zka}.~In the semi-classical calculation one expects a contribution from real pairs that have nucleated plus a contribution from a virtual current.~The former contribution in this limit reproduces~\eref{JlarMexpsmallam}.~However, this is puzzling because based on the understanding in lower dimensions~\cite{Frob:2014zka} one would expect to find~\eref{JlarMexpsmallam} only after adding the virtual
current contribution as well.~We are not able in this work to provide a resolution to this puzzle, but leave further investigation to future work.

\section{Fermion Schwinger current in de Sitter}\label{sec:fermioncurrent}

In this section we present the calculation of the charged fermion Schwinger current in de Sitter space with many details found in~\cite{Hayashinaka:2016qqn}.~While our intermediate result in~\eref{fermiondivcurrentVEV} is in agreement with 
theirs, we differ in the explanation of the several steps involved in computing the complex integrals.~For this reason we provide the details of our derivation in~\aref{Fcurrent} and in this section we summarize the main steps.~As for the scalar current, in the fermion case there are also subtle and important differences with the results obtained in this study and those in the literature due to different renormalization procedures.

We begin with the action for fermion QED,
\begin{equation}\label{Eq:actionFQED}
S=\int d^4 x \sqrt{-g}  \left[-\frac{1}{4}(1+\delta_3)F^{\mu\nu}F_{\mu\nu}-\frac{1}{2} m_A^2 A_\mu A^\mu  +  \overline{\psi}\left(i{\gamma}^\mu(\nabla_\mu + ie A_\mu)-m_\psi \right) \psi \right],
\end{equation}
where $\nabla_\mu$ is the covariant derivative and $m_\psi$ the fermion mass.~Due to the spin 1/2 nature of fermions, the covariant derivative acts as $\nabla_\mu \psi = (\partial_\mu +B_\mu)\psi$, where $B_\mu$ is the spin connection~\cite{Parker:2009uva}.~We use the verbein formalism such that line element can be written as,
\bea
ds^2 = dx^\mu dx^\nu g_{\mu\nu} = w^aw^b\eta_{ab},
\eea
where $\eta_{ab}$ is the Minkowski metric and the basis are related by,
\bea
w^a = e^a_\mu dx^\mu,
\eea
such that $e^a_\mu e^\mu_b = \delta^a_b$ and $e^a_\mu e^\nu_a = \delta^\nu_\mu$.~In this way we can write the spin connection as $B_\mu = \frac{1}{8}w_\mu^{ab}[\gamma_a,\gamma_b]$ where the Dirac matrices are then $\gamma^a = e^a_\mu{\gamma}^\mu$ such that $\{\gamma_a,\gamma_b\} = -2\eta_{ab}$.~Using the de Sitter metric we find for the spin connection $B_0=0$ and $B_i = -\frac{1}{2}\frac{\partial_\tau a}{a^2}\gamma^i\gamma^0$.~With the field redefinition $\xi = a^{3/2}\psi$ we obtain the equations of motion for the charged fermion,
\bea\label{Eq:EOM_spinor}
\left(i\gamma^a\partial_a - eA_a\gamma^a - a m_\psi\right)\xi(\tau)=0 .
\eea
Following~\cite{Hayashinaka:2016qqn} we can make the substitution
$\xi = (i\gamma^a\partial_a-eA_a\gamma^a+m_\psi a)\zeta$
such that the equations of motion for the modified mode functions $\zeta$ can be written as,
\bea
\left(\partial^2_\tau + \omega_k^2(\tau) - i\sigma_s (aH)^2 x \right)\zeta_{s,k}(\tau) = 0 ,
\eea
where $\sigma_{1,2} = +1 $ and $\sigma_{3,4} = -1$, 
\bea
\omega^2_k(\tau) = 
k^2 - 2aH\lambda r k + (aH)^2 x^2 ,
\eea
is the mode frequency and we defined the dimensionless ratios,
\bea\label{Eq:Fratios}
x \equiv \sqrt{\mbpsi^2+\lambda^2},\quad \mbpsi \equiv m_\psi/H , \quad \lambda = eE/H^2 , \quad r = k_z/k \, .
\eea
The solutions are obtained in terms of the Whittaker functions $W_{\kappa,\mu}(z)$ where we take the positive frequency modes in the region $\tau \to - \infty$ which are given by,
\bea
\zeta^+_{s,k} &=& \frac{e^{\pi \lambda r}}{\sqrt{2}k}\sqrt{\frac{x}{x - \lambda r}}
W_{-i\lambda r,1/2+ix}(2ik\tau) . 
\eea
In terms of the positive frequency mode functions, the Schwinger current can be written as,
\bea
\langle J^\psi_z \rangle &\equiv&
\langle 0 | J^\psi_z|0\rangle = -e\langle 0| \overline{\xi}\gamma^3\xi |0\rangle \nn
&=& -\frac{2e\lambda}{x}\int 
\frac{d^3k}{(2\pi)^3}
\Big[1+i\gamma k_z(\zeta^+\zeta^{+*'}-\zeta^{+'}\zeta^{+*}) 
+ 2(F_k^2-\omega_k^2-\gamma F_k\omega_k)|\zeta^+|^2\Big],
\eea
where we have defined $\gamma \equiv x/\lambda -\lambda/x$ and $F_k \equiv \omega_k\omega_k/(aH)^2x$.~Unlike the scalar current, the expectation value of the fermion current also depends on derivatives of the mode functions, making its evaluation more difficult.~As in the scalar case, there are quadratic and logarithmic divergences which can be explicitly seen when introducing a momentum cut-off $P$.~Following the steps outlined in~\aref{Fcurrent} this results in,
\bea\label{Eq:fermiondivcurrentVEV}
\langle J^\psi_z\rangle &=&
- 2aH\frac{e^2 E}{4\pi^2}\lim_{P\rightarrow\infty}
\Big[\frac{2}{3}\Big(\frac{P}{aH}\Big)^2-\frac{2}{3} \ln\Big(\frac{2P}{aH}\Big)
+ \frac{7}{18} - \frac{4\lambda^2}{15} - \frac{\mbpsi^2}{3} \nonumber \\
&-& \frac{3\mbpsi^2}{2\lambda^2}\Big(1+\frac{x}{2\lambda}\ln\Big(\frac{x-\lambda}{x+\lambda}\Big)\Big)-F_\psi(\lambda,\mbpsi)\Big] ,
\eea
where we again have taken $R = 12H^2$ and defined the function $F_\psi(\lambda,\mbpsi)$ as,
\begin{equation}\label{Eq:F_psi}
\begin{aligned}
F_\psi(\lambda,\mbpsi)\equiv & 
\frac{x\text{csch}(2\pi x)}{12\pi^3\lambda^2}\Big\{(45-\pi^2(11-12\lambda^2+8x^2))\cosh(2\pi \lambda)\\
&-(45-\pi^2(11-72\lambda^2+8x^2))\frac{\sinh(2\pi \lambda)}{2\pi \lambda}\Big\}\\
&-\frac{3xM^2\text{csch}(2\pi x)}{8\lambda^3}\sum_{s=\pm}se^{-2\pi xs}(\text{Ei}(2\pi s(x+\lambda))-\text{Ei}(2\pi s(x-\lambda)))\\
&-\frac{\text{csch}(2\pi x)}{4}\operatorname{Re}
\Big[\int_{-1}^1dy(1+x^2 - (1+3\lambda^2 + 3x^2)y^2+5\lambda^2y^4)\\
&\times\sum_{s=\pm}s(e^{2\pi \lambda y} - e^{-2\pi sx})\psi(i(\lambda y + sx))\Big] .
\end{aligned}
\end{equation}
From here on we depart from~\cite{Hayashinaka:2016qqn} which uses adiabatic subtraction whereas we utilize Pauli-Villars regularization following the same procedure as for the scalar case.~We introduce the same three extra heavy fields with the same masses and same combination of wrong kinetic signs as in~\eref{PVcond1}.~Taking the Pauli-Villars mass $\Lambda$ to infinity and dropping terms suppressed by inverse powers of $\Lambda$ we obtain the regularized current,
\bea
\left\langle J^\psi_z\right\rangle_{\mathrm{reg}} 
= 2aH\frac{e^2 E}{4\pi^2} \left[\frac{1}{3}\ln 
\Big(\frac{\Lambda^2}{H^2}\Big)
+\frac{1}{2}+\frac{2 \lambda^2}{15}
+\frac{3\mbpsi^2}{2 \lambda^2}
\left(1+\frac{x}{2\lambda}\ln\left(\frac{x-\lambda}{x+\lambda}\right)\right)+F_\psi(\lambda,\mbpsi)\right] .
~~~~~~
\eea
Finally, the physical renormalized fermion current is then given by,
\bea \label{Eq:fermionJPV_ren}
\langle J_{z }^\psi\rangle_{\rm ren} &=& 
\langle J_{z }^\psi\rangle_{\rm reg} 
+ (2a H E) \delta^\psi_3 \\
&=& eaH^3 \frac{\lambda}{2\pi^2}
\Big[\frac{1}{2} + \frac{2 \lambda ^2}{15}
+ \frac{3 \mbpsi^2}{2 \lambda ^2} 
\Big(1+\frac{x}{2 \lambda } 
\ln (\frac{x -\lambda }{\lambda + x})\Big) 
+ F_\psi(\lambda,\mbpsi) 
+  \frac{1}{3} \ln (\mbpsi^2 + 3)  \nn
&+& \frac{2}{3}(\mbpsi^2 + 3)
- \frac{2}{3} \left( (\mbpsi^2 + 3) - 1\right) \sqrt{2(\mbpsi^2 + 3) + 1} \coth ^{-1}
\Big(\sqrt{2 (\mbpsi^2+3)+1}\Big) -\frac{5}{9}  \Big] , \nonumber
\eea
which as in the scalar case is always \emph{finite and positive}.~Neglecting the curvature correction to the fermion mass (that is taking $\ln(\mbpsi^2 +3) \to \ln \mbpsi^2$), the first five terms match exactly the result found in~\cite{Hayashinaka:2016qqn} which utilized the adiabatic subtraction scheme.~We can reproduce that result by setting $p^2 = 0$ in the vacuum polarization diagram when determining $\delta_3^\psi$.~Doing so leads to a leftover negative IR divergence.~So, as in the scalar case, we see the peculiar negative IR divergence in the renormalized fermion current found in~\cite{Hayashinaka:2016qqn} can be traced to imposing an unphysical renormalization condition with $p^2 = 0$.~As we have emphasized, this is inconsistent with a constant classical electric field in de Sitter space which requires the photon to have a tachyonic mass given by $p^2 = -m_A^2 = 2H^2$.

\subsection{Behavior of the fermion current}\label{sec:fermionlimit}

With the renormalized fermion Schwinger current in hand we can examine its behaviour and study its different limits.~Again, the physical current is defined as $\langle J_z^\psi \rangle_{\rm ren} / a$ allowing us to define a dimensionless (or normalized) fermion current which we will examine below,
\bea\label{Eq:dimlessJpsi}
\mathcal{J}_\psi \equiv  \frac{\langle J_z^\psi \rangle_{\rm ren}}{a e H^3} .
\eea
In~\fref{fermionfig1} we show $\mathcal{J}_\psi$ (solid) as a function of $\lambda$ (top) for various fixed values of $\mbpsi$ and as a function of $\mbpsi$ (bottom) for various fixed values of $\lambda$.~For comparison we also show the conformally coupled scalar (dashed) case presented in the previous section.~We see very similar behavior between the fermion and conformally coupled scalar currents.
\begin{figure}[ht]
	\centering
	\includegraphics[totalheight=9.5cm]{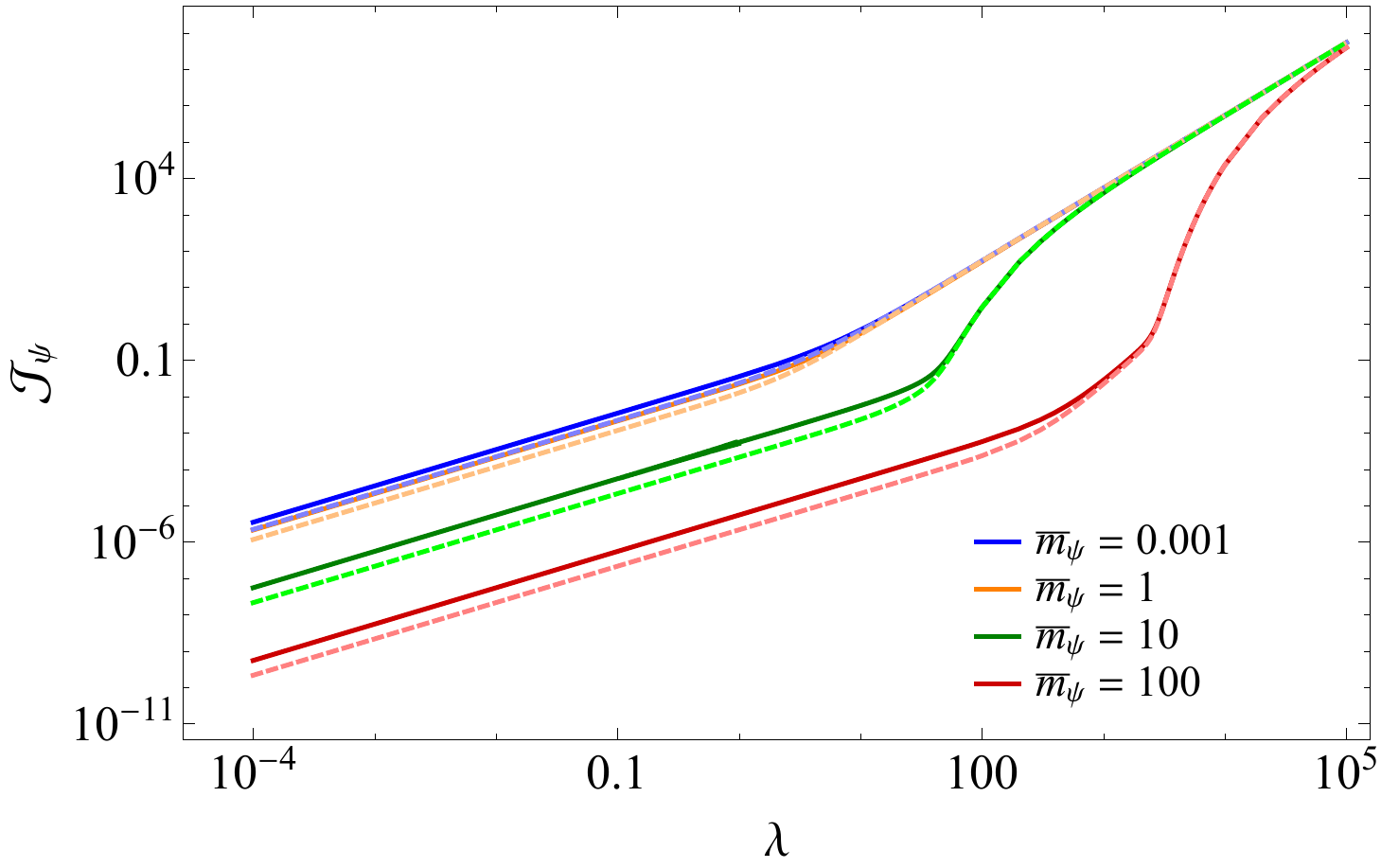} 
	\includegraphics[totalheight=9.5cm]{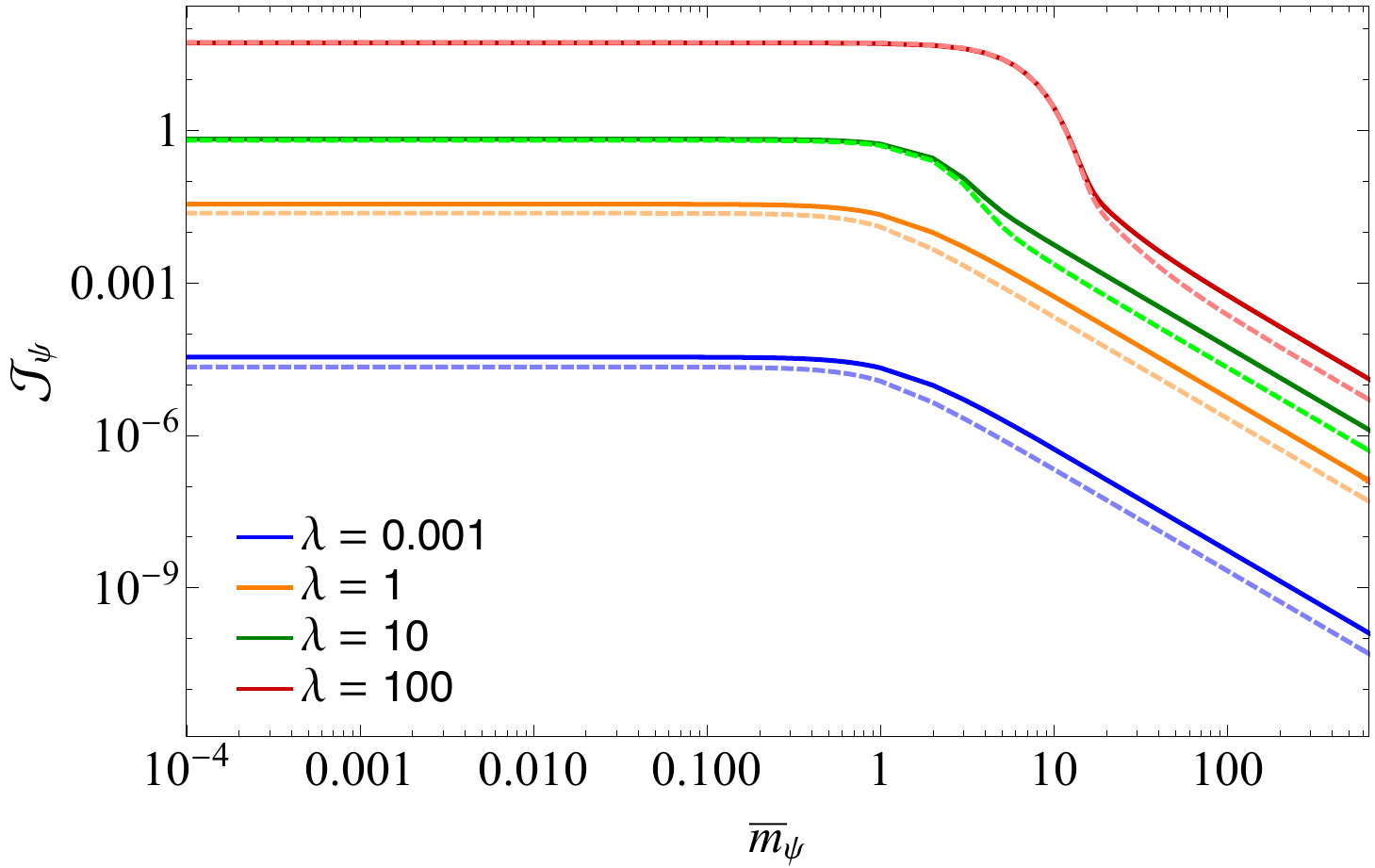}
	\caption{{\bf\,Top:\,}Induced (dimensionless) Schwinger current defined in~\eref{dimlessJpsi} for a charged fermion as a function of $\lambda \equiv eE/H^2$ for various fixed values of $\mbpsi \equiv m_\psi/H$ as shown.\,{\bf Bottom:\,}Same as top, but as a function of $\mbpsi$ for fixed values of $\lambda$ as shown.~For comparison we also show the conformally coupled scalar (dashed) case presented in the previous section.}
	\label{fig:fermionfig1}	
\end{figure}
%


\subsubsection{Strong force limit: $|eE| \gg H^2$}

In the limit $\lambda \gg 1$ the dominant contribution comes from $F_\psi(\lambda , \mbpsi)$ 
in \eref{fermionJPV_ren} giving,
\bea
\mathcal{J}_\psi 
\simeq
\frac{\lambda^2 }{6 \pi^3}
e^{-\pi \mbpsi^2/\lambda} 
\quad (\lambda\gg 1,\mbpsi^2)\, .
\eea
This is twice the result found in the scalar case, as expected, and upon interpreting
$1/H \propto t$ matches to the Schwinger current in flat spacetime.


\subsubsection{Weak force limit: $|eE| \ll H^2$}

In the weak force limit $\lambda \ll 1$ we find,
\bea \label{Eq:fsmallambda}
\mathcal{J}_\psi &\simeq&
\frac{\lambda}{18 \pi ^2}  
\Big[ - \frac{2\pi \mbpsi (1 + 4\mbpsi^2)}{\sinh (2\pi \mbpsi)}
-  \frac{6}{1 + e^{2\pi \mbpsi}} {\rm Re}[e^{2\pi \mbpsi} \psi(- i \mbpsi) + \psi(i \mbpsi) ]  \\
&+& 3 \log \left(\mbpsi^2+ 3 \right)
 - 6 (2 + \mbpsi^2) \sqrt{7+ 2 \mbpsi^2} \coth ^{-1}\left(\sqrt{7+ 2 \mbpsi^2}\right) + 6\mbpsi^2 + 13
    \Big] \, .\nonumber
\eea
The first line comes from expanding $F_\psi(\lambda, \mbpsi)$ while the second from the finite parts of the counterterm.~Taking the massless fermion limit, $\mbpsi \to 0$, of \eref{fsmallambda} we get obtain,
\bea\label{Eq:JFcon}
\mathcal{J}_\psi  \simeq 
\frac{\lambda}{3 \pi^2}\left( (\gamma_E  - \frac{1}{6}) 
+ (\frac{\ln 3}{2} + 3 - 2 \sqrt{7} \coth ^{-1}(\sqrt{7}) - \frac{5}{6}) \right)\,  .
\eea
As for the current for a conformally coupled scalar, we see that in the massless limit the current for a charged fermion again converges to a constant that is fixed by the physical renormalization in~\eref{del3fix2}, but now numerically equal to $\mathcal{J}_\psi \approx 0.035\lambda$ which is about three times larger than the conformally coupled scalar current in the massless limit (see~\eref{JFcon}).~Organizing the terms in the same way as for the conformally coupled scalar in~\eref{JScon}, the two terms in the first parenthesis come from the regularized current while the terms in the second parenthesis come from the counterterm.~Again the first three terms in the second parenthesis arise due to the curvature corrections to the electron mass which are crucial for compensating the negative terms and ensuring a positive renormalized current.~As we can see, the charged fermion current has the same behavior as for a conformally coupled scalar giving us further confidence in our renormalization procedure.


\subsubsection{Large mass limit}

The derivation in this section agrees with and confirms the results obtained in Appendix C of~\cite{Banyeres:2018aax}.~In the large mass limit $\mbpsi \gg 1$ we obtain,
\begin{equation} \label{Eq:JreglarM1f}
\begin{split}
\left\langle J^\psi_z\right\rangle_{\mathrm{ren}} & = \left\langle J^\psi_z\right\rangle_{\mathrm{reg}} 
+ 2 \frac{aH^3}{e} \lambda \delta_3^\psi \\
& =  a e H^3 \frac{\lambda}{2\pi^2}  \left[ - \frac{1}{18} \frac{1}{\mbpsi^2}
+ \frac{1}{\mbpsi^4} \left( -\frac{1}{180} + \frac{2}{45} \lambda^2  \right) 
 + \frac{1}{\mbpsi^6} \left(- \frac{1}{378} +\frac{1}{45} \lambda^2  \right)  \right. \\
& \ \left. \quad \qquad \qquad \quad + \frac{1}{\mbpsi^8} \left( - \frac{1}{360} + \frac{58}{1575} \lambda^2 + \frac{4}{105} \lambda^4  \right) + {\cal O}\left(\frac{1}{\mbpsi^{10}} \right)
\right] \\
& + a e H^3 \frac{\lambda}{2\pi^2} \left[  \frac{17}{15} \frac{1}{\mbpsi^2} 
-  \frac{27}{14} \frac{1}{\mbpsi^4} 
 +  \frac{4139}{945} \frac{1}{\mbpsi^6}  
 -  \frac{22177}{1980} \frac{1}{\mbpsi^8}  + {\cal O}\left(\frac{1}{\mbpsi^{10}} \right)\right] \, .
\end{split}
\end{equation}
The last line in \eref{JreglarM1f} comes from the expansion of the counterterm piece.~As in the scalar case, we expect the terms coming from the regularized piece of the current (first two lines) to match onto the effective currents,
 \bea
 J_{{\rm flat},\mu}^{\rm eff} &=& a e H^3 \frac{\lambda}{2\pi^2} \left( \frac{2}{45}\frac{\lambda^2}{\mbpsi^4} + \frac{4}{105} \frac{\lambda^4}{\mbpsi^8}  \right) + \dots \, , \\ 
J_{{\rm curv},\mu}^{\rm eff} &=& - a e H^3 \frac{\lambda}{2\pi^2} \frac{1}{18 \mbpsi^2}  + \dots \, ,\nonumber
 \eea
which are obtained respectively from the flat space Euler-Heisenberg Lagrangian~\cite{Dunne:2004nc},
\begin{equation} \label{Eq:EHflat}
\begin{split}
{\cal L}_{\rm flat}^{\rm eff} = & \frac{e^4}{90 \pi^2 m_\psi^4} \left[  \left( F_{\mu\nu} F^{\mu\nu} \right)^2 + \frac{7}{4} \left( F_{\mu\nu} \tilde F^{\mu\nu} \right)^2 \right] \\
&  -  \frac{e^6}{630 \pi^2 m_\psi^8} F_{\mu\nu} F^{\mu\nu}
\left[ 8 \left( F_{\mu\nu} F^{\mu\nu} \right)^2 + 13 \left( F_{\mu\nu} \tilde F^{\mu\nu} \right)^2 \right] 
+ {\cal O}\left( \frac{1}{m_\psi^{12}} \right)  \, ,
\end{split}
\end{equation}
and from the curved space Euler-Heisenberg Lagrangian~\cite{Bastianelli:2008cu},
\begin{equation} \label{Eq:EHcurv}
\begin{split}
{\cal L}_{\rm curv}^{\rm eff} =  \frac{e^2}{16 \pi^2 m_\psi^2} & \left[ \frac{1}{36}  R F_{\mu\nu} F^{\mu\nu} - \frac{13}{90} R_{\mu\nu} F^{\mu \alpha} {F^\nu}_\alpha \right. \\
& \left. +\frac{1}{90} R_{\mu\nu\alpha\beta} F^{\mu\nu} F^{\alpha \beta} + \frac{2}{15} \partial^\alpha F_{\alpha\mu} \partial_\beta F^{\beta\mu}  \right] 
+ {\cal O}\left( \frac{1}{m_\psi^{4}} \right)  \, .
\end{split}
\end{equation} 
We do indeed find perfect matching and again we speculate that the terms in 
$\left\langle J^\psi_z\right\rangle_{\mathrm{reg}}$ that do not belong to 
$J_{{\rm flat},\mu}^{\rm eff}$ correspond to the higher order terms 
of the curved Euler-Heisenberg Lagrangian
which are difficult to compute directly~\cite{Bastianelli:2008cu}.~As in the scalar case, besides the terms suppressed by powers of $\mbpsi$ which can be interpreted as the contribution from virtual charged particles, there are also exponentially suppressed terms which might relate to real heavy fermions.~The leading term of the latter is the same as the one in~\eref{JlarMexp} for the scalar and presents the same puzzles which we plan to investigate in a future study.


\section{Summary and conclusions}\label{sec:conclusion}

We have studied constant classical electric fields and the Schwinger effect in de Sitter space which may have implications for inflationary dark matter production as well as magnetogenesis.~We first derived consistency conditions for having a constant electric field in de Sitter space showing that it requires the photon to have a tachyonic mass given by $m_A^2 = -2H^2$ where $H$ is the Hubble scale.~Crucially, in our analysis we have treated the photon field as a classical \emph{dynamical} field, as opposed to a background field.~It must therefore satisfy the Euler Lagrange equations of motion which in turn leads to the tachyonic mass condition.~We then defined the renormalized Lagrangian and a physical renormalization condition (see~\eref{del3fix2}) consistent with a tachyonic photon mass.~This fixes the single counterterm, needed to remove the only divergence present in the theory, which we then computed for the first time (see~\eref{delta3PV} for scalars and~\eref{delta3PVf} for fermions).

With this counterterm we then recomputed the induced Schwinger current in de Sitter for both charged fermions and minimally coupled scalars finding them to be \emph{finite and positive} including in the small mass regime.~In particular, our results are free of the peculiar negative IR divergences found in previous calculations~\cite{Kobayashi:2014zza,Hayashinaka:2016dnt,Hayashinaka:2016qqn}.~In~\cite{Banyeres:2018aax} the authors argued that the negative current is not physical and correctly connected the origin of the issue with the renormalization condition, but did not connect such a condition to the tachyonic mass of the vector, which is a crucial
point of our work.~Here we have examined this issue in detail and traced the origin of these negative IR divergences to renormalization conditions which are inconsistent with the existence of a constant classical electric field in de Sitter space.~We have also computed the Schwinger current in de Sitter space for a conformally coupled scalar again finding it to be positive (in contrast to~\cite{Hayashinaka:2018amz}) and finite with very similar behavior to the charged fermion current.~We then studied the behavior and various limits of the scalar and fermion Schwinger currents (see~\fref{scalarfig1} and~\fref{fermionfig1}).~We find the same IR hyper-conductivity for a light minimally coupled scalar found in previous calculations~\cite{Kobayashi:2014zza,Banyeres:2018aax} and recover the result for the current in flat space in the strong field limit.~However, in the weak field and small mass limits our results differ from previous calculations and in particular, are IR finite.~Since the current is IR safe and always positive, we can consider production of massless (conformally coupled) scalars and fermions during inflation, something which cannot occur from purely gravitational production.

We emphasize that previous calculations of the Schwinger current in de Sitter space~\cite{Kobayashi:2014zza,Hayashinaka:2016qqn,Banyeres:2018aax,Hayashinaka:2018amz}, which treated the electric field strictly as a background field, all base their renormalization procedure on making assumptions about the physical behavior of the current and lead to some combination of negative currents, IR divergences, or different behavior between the conformally coupled scalar and fermion currents.~The renormalization procedure presented here on the other hand relies on assumptions about the electric field as input and treats the photon field as dynamical.~It is thus independent of the current and applies to minimally and non-minimally coupled scalars as well as fermions.~We have shown this leads to a renormalized current which is always finite and positive with virtually identical behavior between the conformally coupled scalar and fermion currents.

Finally, we emphasize that since the gauge field is not quantized (only the charged fields), the one-loop vacuum polarization calculation to fix the counterterm contains the full quantum information so our results are non-perturbative.~While the calculation of the regularized current is exact, the calculation of the counterterm is done with an approximation of the vacuum polarization diagram in de Sitter space, as discussed in~\sref{counterterm}.~Thus our final result for the renormalized current is non-perturbative but not exact and can in principle be improved with a full calculation of the counterterm in de Sitter space.~Nevertheless, our results resolve the puzzling IR behavior found previously in the literature and are an important step in  gaining understanding of the Schwinger effect in de Sitter space.


\vspace{0.3cm}
\noindent
{\bf Acknowledgments:}\,
The authors are especially thankful to Guillem Dom\'{e}nech, Jaume Garriga, and Takeshi Kobayashi for extensive discussions.~The authors also thank Pedro Garcia Osorio, Manel Masip, and Jose Santiago for useful comments and discussions.\,This work has been partially supported by Junta de Andaluc\'ia Projects\, P21-00199, A-FQM-472-UGR20 (fondos FEDER) and by SRA (10.13039/501100011033) and ERDF under grant PID2022-139466NB-C21 (R.V.M.) as well as by PID2022-140831NB-I00 funded by MICIU/AEI/10.13039/501100011033 and FEDER,UE (M.B.G.,A.T.M.),\,FCT CERN grant 10.54499/2024.00252.CERN\,(M.B.G.,A.T.M.,P.F),\,and FCT fellowship SFRH/BD/151475/ 2021 with DOI identifier 10.54499/SFRH/BD/151475/2021 (P.F.).\,LU is supported by the Slovenian Research Agency under the research core funding No.P1-0035, and by the research grants J1-60026 and J1-4389.\,This article is based upon work from COST Action COSMIC WISPers CA21106, supported by COST (European Cooperation in Science and Technology).\,RVM thanks the Mainz Institute for Theoretical Physics (MITP) of the Cluster of Excellence PRISMA$^+$ (Project ID 390831469), for its hospitality and partial support during the completion of this work.

\appendix


\section{Alternative mechanism for constant electric field}
\label{app:conE}

Here we examine an alternative mechanism for generating a constant electric field utilizing a modified kinetic term for a massless photon and show that it is equivalent to our treatment in~\sref{efield} where we used a canonical kinetic term with a tachyonic mass term.~To see this consider the action for a $U(1)$ gauge field, with a modified kinetic term plus a Proca mass term,
\begin{equation}
S = - \int {\rm d}^4 x \sqrt{-g} \left[ \frac{1}{4}I^2(\phi) F_{\mu\nu} F^{\mu\nu} + \frac{1}{2} m_A^2 A_\mu A^\mu \right] \, ,
\end{equation}
with $I(\phi)$ a function of a homogeneous $\phi$, like the inflaton, which only depends on time.~Following Section 2.2 of~\cite{Nakai:2020cfw}, we decompose the vector into transverse and longitudinal modes,
\begin{equation}
A_i = A_i^T + \partial_i \chi \, , \qquad \partial_i A_i^T = 0 \, . 
\end{equation}
Here $\chi$, which we call the longitudinal mode, is related to $A_0$ as follows,
\begin{equation}
A_0 = \frac{-I^2 \partial_i \partial_i}{-I^2 \partial_i \partial_i + a^2 m_A^2} \partial_\tau \chi \, .
\end{equation}
We use $\partial_i = \partial / (\partial x^i)$ for space derivatives, $\partial_\tau$ for conformal time derivatives.~The action splits into transverse and longitudinal,
\begin{equation}
\begin{split}
S & = S_T + S_L \, , \\
S_T & = \frac{1}{2} \int {\rm d}\tau {\rm d}^3x \left[ I^2 \left( \partial_\tau A_i^T \partial_\tau A_i^T - \partial_i A_j^T \partial_i A_j^T   \right) - a^2 m_A^2 A_i^T A_i^T  \right] \, , \\
S_L & =  \frac{1}{2} \int {\rm d}\tau {\rm d}^3x \ a^2 m_A^2 \left[ \partial_\tau \chi \left(  \frac{-I^2 \partial_i \partial_i}{-I^2 \partial_i \partial_i + a^2 m_A^2} \partial_\tau \chi \right) - \partial_i \chi \partial_i \chi \right] \, .
\end{split} 
\end{equation}
In this form it is clear that $S_L = 0$ when (i) $m_A = 0$, or when $m_A \neq 0$ but one considers (ii) a field homogeneous over space $\chi(\vec x, \tau) = \chi(\tau)$, also known as zero mode or infinite wavelength.~In what follows we only consider either case (i) or (ii), so we can forget about the longitudinal mode.

Let us then define the rescaled field,
\begin{equation}
V_i \equiv I(\phi) A_i^T \, .
\end{equation}
After a couple of integrations by parts the action $S_T$ can be brought into the form,
\begin{equation}
S_T  = \frac{1}{2} \int {\rm d}\tau {\rm d}^3x \left[ \partial_\tau V_i \partial_\tau V_i - V_i \left( -\partial_j \partial_j - \frac{\partial_\tau^2 I}{I} + \frac{a^2 m_A^2}{I^2}  \right) V_i   \right] \, ,
\end{equation}
with a canonical kinetic term.~From this we get the equation of motion,
\begin{equation}
\partial_\tau^2 V_i(\vec x, \tau) + \left( -\nabla^2 - \frac{\partial_\tau^2 I}{I} + \frac{a^2 m_A^2}{I^2}  \right) V_i(\vec x, \tau) = 0 \, .
\end{equation}
Now taking a homogeneous field $V_i(\vec x, \tau) = V_i(\tau)$ leads to,
\begin{equation} \label{EOMV}
\partial_\tau^2 V_i(\tau) + \left(- \frac{\partial_\tau^2 I}{I} + \frac{a^2 m_A^2}{I^2}  \right) V_i(\tau) = 0 \, .
\end{equation}
We explicitly consider two scenarios:
\begin{enumerate}
\item $I(\phi) = 1$ and $m_A^2 = -2H^2$;
\item $I(\phi) = a / a_e$ and $m_A^2 = 0$.
\end{enumerate}
As we will see, they both lead to the same homogeneous constant electric field.~Since we are taking the gauge field as homogeneous in space there is no magnetic field.

~\\
\noindent
{\bf Tachyonic mass:}
In the first scenario, $I(\phi) = 1, m_A^2 = -2H^2$, we have $V_i = A_i^T$ and,
\begin{equation}
\partial_\tau^2 V_i(\tau) + \left( -2 a^2 H^2  \right) V_i(\tau) = 0 \, .
\end{equation}
Using $\tau = -1/(aH)$ in de Sitter we have,
\begin{equation} \label{EOMVtac}
\partial_\tau^2 V_i(\tau) - \frac{2}{\tau^2} V_i(\tau) = 0 \, .
\end{equation}
Imposing the boundary condition $V_i(\tau \to -\infty) = 0$ we find the solution,
\begin{equation}
V_i(\tau) = \frac{c}{\tau} \, ,
\end{equation}
with $c$ a constant.~Choosing $c = E/H^2$ and $i = z$, we obtain the gauge field configuration which produces a constant and uniform electric field in de Sitter space.

~\\
\noindent
{\bf Modified kinetic term:}
In the second scenario, we don't introduce a tachyonic mass taking $m_A = 0$ and instead we modify the kinetic term with,
\begin{equation}
I(\phi) = \frac{a}{a_e} = \frac{\tau_e}{\tau} \, ,
\end{equation}
where $a_e$ and $\tau_e$ are constants which could be fixed as the time (or scale factor) at the end of inflation or anything else.~With this choice we have,
\begin{equation}
\frac{\partial_\tau^2 I}{I} = \frac{2}{\tau^2} \, .
\end{equation}
It follows that \eqref{EOMV} reduces to \eqref{EOMVtac} so we have exactly the same equation of motion as in the case above with the tachyonic mass.~Thus we have another way to generate a constant electric field.~Here $I(\phi)$ is what sources the electric field and is proportional to $a = e^{Ht}$, the exponential enhancement needed to compensate for the de Sitter expansion and keep the electric field constant. 

~\\
\noindent
{\bf Constant electric field:}
In both scenarios the electric field originates from a transverse component of the gauge field, there is no contribution from the longitudinal mode.~The gauge field is treated purely as a classical field, we never need to mention quantization.~The comoving electric field is given by $E_\mu = u^\nu F_{\mu\nu}$, with $u^\nu = (1/a,0,0,0)$ the comoving observer 
velocity satisfying $g_{\mu\nu} u^\mu u^\nu = -1$ and $u^i = 0$.~Then $E_\tau = 0$ and we only have $E_i$.~In both scenarios the comoving electric field must be defined in terms of the canonically normalized field $V_i$ as,
\begin{equation}
E_i = u^0 F_{i0}^{\rm canonical} = - \frac{1}{a} \partial_\tau V_i(\tau) \, . 
\end{equation}
Given $V_i(\tau) = E/(H^2 \tau) \delta_i^z$ we get $E_i = a E \delta_i^z$ and the constant field strength,
\begin{equation}
g^{ij}E_i E_j = \frac{1}{a^2} (a E)(a E) = E^2 \, .
\end{equation}

\section{Calculation of the QED Counterterm ($\delta_3$)}
\label{app:delta3}

Here we present details of the calculation of $\delta_3$ obtained by computing the vacuum polarization diagram (see~\fref{selfenergy}) for the photon 2-point function following Ch.16.2 of~\cite{Schwartz:2014sze}.

\subsection{Vacuum polarization in scalar QED}
\label{app:delta3S}

Starting with the vacuum polarization in flat space due to a charged scalar and converting the expressions in chapter 16.2 of~\cite{Schwartz:2014sze} to our metric signature we find,
\begin{equation}
	\Pi _2^{\mu \nu }=2i\int \frac{d^4 k}{(2 \pi )^4}\int _0^1 dx\frac{ \left(g^{\mu \nu } \left(k^2+M_\phi^2+p^2 x^2\right)-2 k^{\mu } k^{\nu }\right)}{\left(k^2-\Delta ^2\right)^2}\,,
\end{equation}
where $\Delta^2 = -M_\phi^2-x(1-x)p^2$ and $M_\phi^2 = m_\phi^2 + 12\xi H^2$.  Taking $k^\mu k^\nu\rightarrow1 /4 k^2 g^{\mu\nu}$ and with a Wick rotation $k^0\rightarrow i k_E$ we obtain,  
\begin{equation}
	\Pi _2^{\mu \nu }=-2e^2g^{\mu \nu }\frac{1}{8 \pi ^2}\int_0^1 dx\int_0^{\infty }dk_E\, k_E^3\frac{ \left(-\frac{k_E^2}{2}+M_\phi^2+p^2 x^2\right)}{\left(\Delta ^2+k_E^2\right)^2}\,.
\end{equation}
To be consistent with the regularization of the current we regularize with the same Pauli-Villars fields where $m_0 = M_\phi$ is the original field and $m_1^2 = m_3^2 = 2\Lambda^2$ and $m_2^2 = 4\Lambda^2 - M_\phi^2$ such that,
\begin{align}
	\tilde{\Pi}^{\mu\nu}_2(m^2,p^2,\Lambda^2) = \sum_{i=0}^3(-1)^i\Pi^{\mu\nu}_2(p^2,m_i^2) ,
\end{align}
which leads to,
\begin{equation}
	\begin{aligned}
		\Pi _2^{\mu \nu }=-2e^2g^{\mu \nu }\frac{1}{8 \pi ^2}\int _0^1dx\int _0^{\infty }dk_E\, k_E^3 \left(\frac{-\frac{k_E^2}{2}+M_\phi^2+p^2 x^2}{\left(\Delta ^2+k_E^2\right)^2}-\frac{-\frac{k_E^2}{2}+m_1^2+p^2 x^2}{\left(\Delta _1^2+k_E^2\right)^2}\right. \\ \left.+\frac{-\frac{k_E^2}{2}+m_2^2+p^2 x^2}{\left(\Delta _2^2+k_E^2\right)^2}-\frac{-\frac{k_E^2}{2}+m_3^2+p^2 x^2}{\left(\Delta _3^2+k_E^2\right)^2}\right)\, .
	\end{aligned}
\end{equation}
Doing the integration in the momentum $k_E$ we find,
\bea
\Pi _2^{\mu \nu }=-e^2 \left(-p^2 g^{\mu \nu }\right)\frac{1}{8 \pi ^2}\int _0^1 dx\  x (2 x-1) \ln \left(\frac{\left(p^2 (x-1) x-M_\phi^2\right) \left(M_\phi^2+p^2 (x-1) x-4 \Lambda ^2\right)}{\left(p^2 (x-1) x-2 \Lambda ^2\right)^2}\right)\,.~~~~\,
\eea
The argument of the logarithm can be simplified under the assumption that the regulator $\Lambda\rightarrow \infty$.~Keeping the leading term we find,
\begin{equation}
	\begin{aligned}
		\Pi _2^{\mu \nu }	&=-e^2 \left(-p^2 g^{\mu \nu }\right)\frac{1}{8 \pi ^2}\int _0^1 dx\  x (2 x-1) \ln \left(\frac{M_\phi^2-p^2 (x-1) x}{\Lambda ^2}\right)\\
		&=-e^2  \left(-p^2 g^{\mu \nu }\right)\frac{1}{144 \pi ^2 p^3}\left(3 p^3 \log \left(\frac{M_\phi^2}{\Lambda ^2}\right)\right.\\&\left.+6 \left(4 M_\phi^2+p^2\right)^{3/2} \tanh ^{-1}\left(\frac{p}{\sqrt{4 M_\phi^2+p^2}}\right)-24 M_\phi^2 p-8 p^3\right)\,.
	\end{aligned}
\end{equation}


\subsection{Vacuum polarization in fermionic QED}
\label{app:delta3F}

In QED, there is only one diagram at first order that contributes to the photon propagator, see Ch.\,16.2 of \cite{Schwartz:2014sze} that with our metric signature gives,
\begin{align}
	\Pi_2^{\mu\nu} = 4ie^2\int \frac{d^4k}{(2\pi)^4}\int_0^1dx\frac{2k^\mu k^\nu-g^{\mu\nu}[k^2-x(1-x)p^2+\bar{m}_f^2]}{[k^2-\Delta^2]^2},
\end{align}
after summing over the Dirac matrices and $\Delta^2$ is defined in a similar way as for the scalars substituting $M_\phi \rightarrow M_\psi$ where $M_\psi^2 = m_\psi^2 + 3H^2$.~Following the same procedure as above as for the scalar case taking $k^\mu k^\nu \rightarrow k^2/4g^{\mu\nu}$ and $k^0\rightarrow ik_E$ gives,
\begin{align}
	\Pi_2^{\mu\nu} = -\frac{4}{8\pi^2}e^2g^{\mu\nu}\int dk_E k_E^3\int_0^1dx\frac{-\frac{k_E^2}{2}+x(1-x)p^2-\bar{m}_f^2}{[-k_E^2+\Delta^2]^2}.
\end{align}
Again utilizing Pauli-Villars regularization with the same three fields as in the scalar case we perform the integral in $k_E$ and expanding for large $\Lambda$ we obtain,
\begin{align}
	\tilde{\Pi}^{\mu\nu}_2 &= \frac{4}{8\pi^2}e^2(-p^2)g^{\mu\nu}\int_0^1 dxx(x-1)\log\Big(\frac{M_\psi^2-p^2x(x-1)}{\Lambda^2}\Big)\\
	&=\frac{e^2}{36\pi^2}(-p^2)g^{\mu\nu}\Big[5-\frac{12M_\psi^2}{p^2}-6\frac{(p^2-2M_\psi^2)\sqrt{4M_\psi^2+p^2}}{p^3}\coth ^{-1}\frac{\sqrt{4M_\psi^2+p^2}}{p}-3\log\frac{M_\psi^2}{\Lambda^2}\Big] .\nonumber
\end{align}
Finally, noticing that $\Pi^{\mu\nu} = -e^2p^2g^{\mu\nu}\Pi_2(p^2)$ and imposing the renormalization condition in~\eref{del3fix2} we obtain the counter term (defining $\Mbpsi = M_\psi /H$),
\bea\label{Eq:delta3PVfapp}
\delta_3^{\psi} &=& 
 -e^2\Pi(2H^2) \\
 &=& 
\frac{e^2}{36\pi^2 }
\Big[
-3 \ln \left(\frac{\Lambda ^2}{H^2}\right)
+ 3 \ln (\Mbpsi^2) 
- 6 (\Mbpsi^2 - 1) 
\sqrt{2\Mbpsi^2 +1}\, \coth ^{-1}
\left({\sqrt{2\Mbpsi^2 + 1}}\right) 
+ 6 \Mbpsi^2  - 5\Big] . \nonumber
\eea
Again this is free of IR divergences in the $\Mbpsi \to 0$ limit which gives,
\bea\label{Eq:delta3PVfapplim}
\delta_3^{\psi} &=& 
 -e^2\Pi(2H^2) \\
 &=& 
\frac{e^2}{36\pi^2 }
\Big[
-3 \ln \left(\frac{\Lambda ^2}{H^2}\right)
+ \ln8  - 5\Big] . \nonumber
\eea
%

\section{Calculation of the fermion current}
\label{app:Fcurrent}

Starting with the definitions in~ \sref{fermioncurrent}, in this appendix we review the calculation of the fermionic Schwinger current.~Following~\cite{Hayashinaka:2016qqn} the solutions of the Dirac equation can be written as,
\begin{align}
	\xi = (i\gamma^a\partial_a-eA_a\gamma^a+m_\psi a)\zeta,
\end{align}
such that we have,
\begin{align}
	(\partial_\tau^2+\omega_k(\tau)^2-i\sigma_s\sigma(\tau))\zeta_{s,k}(\tau)=0 ,
	\label{Eq:doubledirac}
\end{align}
where $\sigma_{1,2} = +1 $ and $\sigma_{3,4}=-1$ are eigenvalues of the matrix $(\overline{m}^2_\psi+\lambda^2)^{-1/2}(\overline{m}_\psi\gamma^0+\lambda\gamma^0\gamma^3)$.~The time-dependent frequency is given by,
\begin{align}
	\omega^2_k(\tau) = k^2-2aH\lambda k_z+(aH)^2(\overline{m}^2_\psi+\lambda^2),\quad \sigma(\tau) = (aH)^2\sqrt{\overline{m}^2_\psi+\lambda^2}.
\end{align}
The solutions of~\eref{doubledirac} are given by the Whittaker function $W_{\kappa,\mu}(z)$ with parameters,
\begin{align}
	&\zeta^+ = \frac{e^{\pi \lambda r}}{\sqrt{2}k}\sqrt{\frac{x}{x-\lambda r}}W_{-i\lambda r,1/2+ix}(2ik\tau)\\
	&\kappa = -i\lambda \frac{k_z}{k} = -i\lambda r,\quad \mu_s = \frac{1}{2}-i\sigma_s\sqrt{\overline{m}^2_\psi+\lambda^2} = \frac{1}{2}-\lambda_sx,\quad z = 2ik\tau,
\end{align}
where we define $x=\sqrt{\overline{m}^2_\psi+\lambda^2}$.~The current can be further simplified to, using the normalisation condition $\zeta'(\zeta^*)'-iF_k(\zeta(\zeta^*)'-\zeta'\zeta^*)+\omega_k^2|\zeta|^2=0$,
\begin{align}
	\langle J^3\rangle = -\frac{2e\lambda}{r}\int \frac{d^3k}{(2\pi)^3}\Big[1+i\gamma k_z(\zeta^+\zeta^{+*'}-\zeta^{+'}\zeta^{+*})+2(F_k^2-\omega_k^2-\gamma F_k\omega_k)|\zeta^+|^2\Big],
\end{align}
with $\gamma =x/\lambda -\lambda/x$ and $F_k = \omega_k\omega_k'/\sigma$. 

Using a cut off momentum $P$ as a regulator initially, the first term of the current is,
\begin{align}
	-\frac{2e\lambda}{x}\int_0^{\infty} \frac{d^3k}{(2\pi)^3}= -\lim_{P \rightarrow \infty}\frac{P^3}{6\pi^2 x} .
\end{align}
For the other terms of the current, we also consider a cut-off regulator and write the Whittaker functions in the Milles-Barnes integral representation as in~\cite{Kobayashi:2014zza},
\begin{align}
	W_{\kappa,\mu}(z) = \int_{C_s}\frac{ds}{2\pi i}z^{-s}e^{-z/2}\frac{\Gamma(-s-\kappa)\Gamma(s-\mu+1/2)\Gamma(s+\mu-1/2)}{\Gamma(1/2-\kappa-\mu)\Gamma(1/2-\kappa+\mu)}.
\end{align}
The integral is performed along a path $C_s$ that goes from $-i\infty$ to $+i\infty$ and separates the poles from $\Gamma(-s-\mu)$ and $\Gamma(s-\mu+1/2)\Gamma(s+\mu-1/2)$.~From now on, in the full computation of the current, we follow the conventions on \cite{Kobayashi:2014zza}.~Using the properties of the Whittaker functions $(W_{\kappa,\mu}(z))^* = W_{\kappa^*,\mu^*}(z^*)$ and $dW_{\kappa,\mu}(z)/dz =(1/2-\kappa/z)W_{\kappa,\mu}(z)-W_{1+\kappa,\mu}(z)/z$, the current can be further written as,
\begin{align}
	&-2e\lambda \lim_{P\rightarrow \infty}\int_0^P dk\int_{-1}^1dr\int_{C_s}\frac{ds}{2\pi i}\int_{C_t}\frac{dt}{2\pi i} \frac{e^{\pi \lambda r}}{4\pi^2}e^{i\frac{\pi}{2}(s-t)}\Gamma(-s+i\lambda r)\Gamma(s-ix)\Gamma(s+ix+1)\nonumber\\
	&\times \Gamma(-t-i\lambda r)\Gamma(t+ix)\Gamma(t-ix+1)\frac{\sinh\pi(x-\lambda r)\sinh\pi(x+\lambda r)}{\pi^2(x+\lambda r)}\Big(\frac{k}{aH}\Big)^{-s-t}\nonumber\\
	&\times\Big[(r^2+\gamma r -1)k^2-aH\gamma(x+\lambda r)(1+\frac{1}{2}\Big(\frac{1+i(x-\lambda r)}{-s+i\lambda r-1}+\frac{1-i(x-\lambda r)}{-t-i\lambda r-1}\Big))kr\Big].
\end{align}
The integration in $k$ is done considering that the contour satisfies $\text{Re}(s), \text{Re}(t)<0$ such that, once performed the integration,
\begin{align}
	&-2e\lambda(aH)^3 \lim_{P\rightarrow \infty}\int_{-1}^1dr\int_{C_s}\frac{ds}{2\pi i}\int_{C_t}\frac{dt}{2\pi i} \frac{e^{\pi \lambda r}}{4\pi^2}e^{i\frac{\pi}{2}(s-t)}\Gamma(-s+i\lambda r)\Gamma(s-ix)\Gamma(s+ix+1)\nonumber\\
	&\times \Gamma(-t-i\lambda r)\Gamma(t+ix)\Gamma(t-ix+1)\frac{\sinh\pi(x-\lambda r)\sinh\pi(x+\lambda r)}{\pi^2(x+\lambda r)}\\
	&\times\Big[(r^2+\gamma r -1)\frac{P^{3-s-t}}{-s-t+3}-aH\gamma(x+\lambda r)(1+\frac{1}{2}\Big(\frac{1+i(x-\lambda r)}{-s+i\lambda r-1}+\frac{1-i(x-\lambda r)}{-t-i\lambda r-1}\Big))r\frac{P^{2-s-t}}{-s-t+2}\Big].\nonumber
	\label{Eq:fermioncurrent1}
\end{align}
In contrast to the scalar current~\cite{Kobayashi:2014zza}, due to the derivatives of the Whittaker functions, the fermionic current has two extra poles\,\footnote{The two extra poles were not mentioned in the calculation of the un-regulated fermion current performed in~\cite{Hayashinaka:2016qqn} though we are in agreement with their final result.}
$s = i\lambda r-1$ and $t = i\lambda r+1$ coming from the last term of~\eref{fermioncurrent1}.~The poles coming from the gamma functions $s = i\lambda r+n$ are on the right side of the contour $C_s$ and the poles $s = ix-n$, $s = -ix-1-n$ are on the left side of the same contour.~In the case of the contour $C_t$, the poles $t = -i\lambda r+n-1$, $t=-s+2$ and $t=-s+3$ are on the right side and $t = -ix-n$ and $t=ix-1-n$ are on the left side.

We start by performing the integration in $t$ choosing a path $C_s$ that satisfies  $-1<\text{Re}(s)$ and closing the path $C_t$ on the right side without passing through any of the poles.~The extra contribution does not affect the final result since it vanishes for an integration along a real part over $t$ and also because in the limit of $\xi\rightarrow \infty$ any contour with $Re(t)>4$ vanishes.~The integral can be written in a power series of $\mathcal{O}(P^{3-s-t})$ and $\mathcal{O}(P^{2-s-t})$ such that the only non-vanishing contribution comes from the poles,
\begin{align}
	t=-i\lambda r-1,-i\lambda r,-i\lambda r+1,-i\lambda r+2,-i\lambda r+3,3-s,2-s ,
\end{align}
when taking $P\rightarrow \infty$.~The result after integrating in $t$ can be separated into,
\begin{align}
	\mathcal{O}(P^{0})+\mathcal{O}(P^{4+i\lambda r -s},...,P^{i\lambda r-s}) ,
	\label{Eq:sep1}
\end{align}
where the contribution $\mathcal{O}(P^0)$ comes from the poles $3-s$ and $2-s$.~Considering the powers dependent on $s$, we close the contour $C_s$ on the right-hand side where a non-vanishing contribution comes from the poles:
\begin{align}
	s = i\lambda r-1,i\lambda r, i\lambda r +1,i\lambda r+2,i\lambda r+3, i\lambda r +4.
\end{align}
Notice that we included the pole $s = i\lambda r +4$ as needed.~The result can be separated again into,
\begin{align}
	-\frac{P^3}{6\pi^2 r}+(aH)\frac{P^2}{6\pi^2}-\frac{(aH)^3}{6\pi^2}\log\frac{P}{aH}+(aH)^3\times\mathcal{O}(P0).
	\label{Eq:sep2}
\end{align}
The zeroth order of~\eref{sep2} is given by,
\begin{align}
	&\frac{\gamma_E}{6\pi^2}-\frac{23}{144\pi^2}-\frac{7\lambda^2}{120\pi^2}+\frac{\lambda^4}{420\pi^2}-\frac{23}{4\pi^2}\frac{\overline{m}^2_\psi}{\lambda^2}+\frac{\lambda^2 \overline{m}^2_\psi}{1440\pi^2}-\frac{5\overline{m}_\psi^4}{576\pi^2}+i\Big(\frac{1}{12\pi}-\frac{1}{2\pi^2 x}+\frac{121\lambda^2}{216\pi^2 x}\nonumber\\
	&-\frac{91\lambda^4}{1440\pi^2 x}+\frac{\lambda^6}{2016\pi^2 x}-\frac{65x}{144\pi^2}+\frac{89\lambda^2 x}{1440\pi^2}-\frac{\lambda^4x}{1120\pi^2}-\frac{41\overline{m}_\psi^2 x}{576\pi^2}-\frac{\lambda^2\overline{m}^2_\psi x}{1440\pi^2}-\frac{\overline{m}^4_\psi x}{576\pi^2}\Big)+\\
	&\frac{1}{8\pi^2}\int_{-1}^1 dr(1+x^2-(1+3\lambda^2+3x^2)r^2+5\lambda^2r^4)(\psi(i\lambda r-ix)+\psi(i\lambda r+ix))-\frac{3\overline{m}^2_\psi x}{8\pi^2\lambda^3}\log\frac{x-\lambda}{x+\lambda}, \nonumber
\end{align}
with $\gamma_E = 0.577216$ is the Euler constant and $\psi$ is the digamma function. 

The zero order of~\eref{sep1} still to be integrated in $s$ is, 
\begin{align}
	\int_{-1}^1drh(r)\Big(\int\frac{ds}{2\pi i}e^{i\pi s}\frac{f(s)}{\sin\pi(s-i\lambda r)\sin\pi(s-ix)\sin\pi(s+ix)}\Big),
	\label{Eq:zero1}
\end{align}
where $h(r) = e^{\lambda r}\sinh\pi(r-\lambda r)\sinh\pi(r+\lambda r)$ and $f(s) = g(s)-g(s-1)+\frac{d}{s-i\lambda r+1}$ where,
\begin{align}
	g(s) = b_3s^3+b_2s^2+b_1s+\frac{c_0}{s-i\lambda r}+\frac{c_1}{s-i\lambda r-1}+\frac{c_2}{s-i\lambda r-2}+\frac{c_3}{s-i\lambda r-3} .
\end{align}
Due to the periodicity of the integrand function, one may shift $s\rightarrow s+1$ such that,
\begin{align}
	\int dr\int \frac{ds}{2\pi i}(...)f(s) = \int dr\Big(\int_{C_s} -\int_{C_{s-1}}\Big)\frac{ds}{2\pi i}(...)g(s)+\int dr\int_{C_s}\frac{ds}{2\pi i}(...)\frac{d}{s-i\lambda r+1}.
\end{align}
where $(...)$ represents the integrand of \eref{zero1}.
The integral in $d$ will contain the infinite sum of the poles $s=i\lambda r-1-n,-ix-n,ix-1-n$ for $n=0,1,2...$ leading to \cite{Hayashinaka:2016qqn},
\begin{align}
	\int_{C_s}\frac{ds}{2\pi i}(...)\frac{d}{s-i\lambda r+1} &\sim d\times \Big[\frac{e^{\pi( \lambda r- x)}\sinh\pi(x+\lambda r)}{\sinh2\pi x}\Big(\gamma_E+\psi(-ix+i\lambda r-1)\Big)+\nonumber\\
	&+\frac{e^{\pi( \lambda r+x)}\sinh\pi(x-\lambda r)}{\sinh2\pi x}\Big(\gamma_E+\psi(i\lambda r+ix)\Big)\Big],
	\label{Eq:intind1}
\end{align}
where $d = \frac{1}{8\pi^2}(1+x^2-(1+3\lambda^2+3x^2)r^2+5\lambda^2r^4)$.~The last equation still needs to be integrated in $r$ and to do so we used the property $\psi(z-1) = \psi(z)-\frac{1}{z-1}$ such that,
\begin{align}
\psi(-ix+i\lambda r-1) = \psi(-ix+i\lambda r)-\frac{1}{-ix+i\lambda r-1}.
\label{Eq:approxford}
\end{align}
Moreover, when computing the integral in \eref{intind1} after including \eref{approxford}, we found terms of the form $\text{Li}_n(e^{2\pi iy})$ and $\text{Li}_n(e^{-2\pi iy})$ where $y$ is a linear combination of $\lambda$ and $x$ and $\text{Li}_n(z)$ are the polylogarithm functions.~These can be written in terms of the Bernoulli polynomials $B_n(x)$ as \cite{abramowitz1964handbook}
\begin{align}
	&\text{Li}_n(e^{2\pi iy}) +(-1)^n \text{Li}_n(e^{-2\pi iy}) = -\frac{(2\pi i)^n}{n!}B_n(x) .
\end{align}
Finally, the integral on $g(s)$ can be computed by considering the poles $s = i\lambda r-1,-ix,ix+1$ inside the closed contour generated by $C_s$ and $C_{s-1}$.~Summing up everything it gives the un-renormalised fermion current in terms of the momentum cut-off $P$, 
\begin{align}
	&\langle J^3\rangle = -2e\lambda (aH)^3\lim_{P \rightarrow \infty}\Bigg[\frac{1}{6\pi^2}\Big(\frac{P}{aH}\Big)^2-\frac{1}{6\pi^2}\log\frac{2P}{aH}+\frac{7}{72\pi^2}-\frac{\lambda^2}{15\pi^2}-\frac{\overline{m}^2_\psi}{12\pi^2}\\
	&-\frac{3\overline{m}^2_\psi}{8\pi^2\lambda^2}\Big(1+\frac{x}{2\lambda}\log\frac{x-\lambda}{x+\lambda}\Big)+\frac{3x\overline{m}^2_\psi\text{csch}(2\pi x)}{32\pi^2\lambda^3}\sum_{s=\pm}se^{-2\pi x s}(\text{Ei}(2\pi s(x+\lambda))-\text{Ei}(2\pi s(x-\lambda)))\nonumber\\
	&-\frac{x\text{csch}(2\pi x)}{48\pi^5\lambda^2}\Big((45-\pi^2(11-12\lambda^2+8x^2))\cosh(2\pi \lambda)-(45-\pi^2(11-72\lambda^2+8x^2))\frac{\sin(2\pi\lambda)}{2\pi \lambda}\Big)\nonumber\\
	&+\frac{\text{csch}(2\pi x)}{16\pi^2}\int_{-1}^1dr(1+x^2-(1+3\lambda^2+3x^2)r^2+5\lambda^2r^4)
	\sum_{s=\pm}s(e^{2\pi\lambda r}-e^{-2\pi x s})\text{Re}[\psi(i(\lambda r+xs))]\Bigg].\nonumber
\end{align}

\bibliographystyle{JHEP}
\bibliography{SchwingerCurrent.bib}

\end{document}